\documentclass{emulateapj}

\usepackage{subfigure}
 \usepackage{graphicx}
\usepackage{threeparttable}
\usepackage{rotating}
\usepackage{lipsum}
\usepackage{arydshln}
\usepackage{amsmath}

\newcommand{\lya}{ Lyman-$\alpha \;$}

  \newcommand{\ergscm}{$\rm erg \; s^{-1}\; cm^{-2}$}
    
    \newcommand{\ergscma}{$\rm erg \; s^{-1}\; cm^{-2}\; \AA^{-1}$}

\newcommand{\gp}{$g^\prime$}
\newcommand{\rp}{$r^\prime$}
\newcommand{\zp}{$z^\prime$}
\newcommand{\zs}{$z\sim7$}
   
    \newcommand{\msun}{$\rm M_{\odot}$}
        \newcommand{\msol}{$\rm M_{\odot}$}
        \newcommand{\msunyear}{$\rm M_{\odot}$ yr$^{-1}$}        
     \newcommand{\spa}{$ \;\;\;\;$}
     
          \newcommand{\psix}{$\mathcal P_6$}
 \newcommand{\zphot}{$z_{\rm phot}$}

 \newcommand{\jone}{\textit {J$_1$}} 
 \newcommand{\jtwo}{\textit {J$_2$}} 
  \newcommand{\jthree}{\textit {J$_3$}} 
   \newcommand{\hs}{\textit {H$_s$}} 
    \newcommand{\hl}{\textit {H$_l$}} 
     \newcommand{\ks}{\textit {K$_s$}}

   \newcommand{\yhst}{\textit {Y$_{105}$}}
  \newcommand{\jhst}{\textit {J$_{125}$}}
   \newcommand{\hhst}{\textit {H$_{160}$}} 
   \newcommand{\muv}{$M_{UV}$}

\newcommand{\spitzer}{\textit{Spitzer}}
\newcommand{\hst}{\textit{HST}}

 \newcommand{\FS}{FourStar }
\newcommand{\zf}{zFourGE }
  
   \newcommand{\zsol}{$\rm Z_{\odot}$}

\newcommand{\bb}{broad-band }
\newcommand{\mb}{medium-band }
 \newcommand{\td}{T-dwarf }
 \newcommand{\nir}{near-IR }
 
\slugcomment{}

\shorttitle{Discovery of Lyman Break Galaxies  at $ z\sim7$ from the \zf Survey}
\shortauthors{Tilvi et al.}

\begin{document}

\title{Discovery of Lyman Break Galaxies  at $ z\sim7$ from the ZFOURGE  Survey
}

\author{V.  Tilvi  \altaffilmark{1},
C. Papovich\altaffilmark{1},
K.-V. H. Tran\altaffilmark{1},
I. Labb{\'e}\altaffilmark{2},
L. R. Spitler\altaffilmark{3},
C. M. S. Straatman\altaffilmark{2},
S. E. Persson\altaffilmark{4},
A. Monson\altaffilmark{4},
K. Glazebrook\altaffilmark{3},
R. F. Quadri\altaffilmark{4,11},
P. van Dokkum\altaffilmark{5},
M. L. N. Ashby\altaffilmark{6}, 
S. M. Faber\altaffilmark{7}, 
G. G. Fazio\altaffilmark{6}
S. L. Finkelstein\altaffilmark{8,11}
H. C. Ferguson\altaffilmark{9}, 
N. A. Grogin\altaffilmark{9}, 
G. G. Kacprzak\altaffilmark{3,12},
D. D. Kelson\altaffilmark{4}, 
A. M. Koekemoer\altaffilmark{9},
D. Murphy\altaffilmark{4}, 
P. J. McCarthy\altaffilmark{4},
J. A. Newman\altaffilmark{10}
B. Salmon\altaffilmark{1},
S. P. Willner\altaffilmark{6}
 }

\altaffiltext{1}{George P. and Cynthia Woods Mitchell Institute for Fundamental Physics and Astronomy, and Department of Physics and Astronomy, Texas A\&M University, College Station, TX. }
\altaffiltext{2}{ Sterrewacht Leiden, Leiden University, NL-2300 RA Leiden, The Netherlands}
\altaffiltext{3}{ Centre for Astrophysics \& Supercomputing, Swinburne University, Hawthorn, VIC 3122, Australia}
\altaffiltext{4}{ Carnegie Observatories, Pasadena, CA 91101, USA}
 \altaffiltext{5}{ Department of Astronomy, Yale University, New Haven, CT 06520, USA}
 \altaffiltext{6}{ Harvard-Smithsonian Center for Astrophysics, 60 Garden St., Cambridge, MA 02138 USA}
   \altaffiltext{7}{UCO/Lick Observatory, Department of Astronomy and Astrophysics, University of California, Santa Cruz, CA, USA}
  \altaffiltext{8}{University of Texas, Austin, TX, USA}  
  \altaffiltext{9}{Space Telescope Science Institute, Baltimore, MD, USA}  
  \altaffiltext{10}{University of Pittsburgh, PA, USA}  
 \altaffiltext{11}{ Hubble Fellow}
\altaffiltext{12}{Australian Research Council Super Science Fellow}

\begin{abstract}

Star-forming galaxies at redshifts $z > 6$ are likely responsible for the reionization of the universe, and it is important to  study
 the nature of these galaxies. We present three candidates for $z \sim 7$ Lyman$-$break galaxies (LBGs) from a 155 
 arcmin$^2$ area in the CANDELS/COSMOS field imaged by the deep \FS  Galaxy Evolution  (\zf ) survey.
The   \FS \mb  filters provide the equivalent of $R\sim10$ spectroscopy, which cleanly distinguishes between 
\zs\ LBGs and brown dwarf stars. 
The distinction between stars and galaxies based on an object's  angular size can become unreliable even when using HST imaging; 
there exists at least one very compact \zs\ candidate (FWHM$\sim $0.5-1 kpc) that is indistinguishable from a point source.
%
%the size of one of our bright \zs\   candidates is indistinguishable from a star. Moreover, there exists a population of very compact 
%(FWHM$\sim $0.5-1 kpc)  galaxies at \zs.
The \mb filters provide   narrower redshift distributions compared  with  broad-band-derived 
redshifts. 
The UV luminosity function derived using the three $z \sim 7$ candidates is consistent with previous studies, suggesting an
 evolution at the bright end ($M_{UV} \sim -21.6$ mag) from $z\sim7$ to $z\sim5$. 
Fitting the galaxies' spectral energy distributions,  we predict \lya\ equivalent widths for the two brightest LBGs, and 
 find that the presence of a  \lya\  line affects the \mb  flux thereby changing the constraints on
 stellar masses and UV spectral slopes.
This illustrates the  limitations of deriving LBG properties  using only  \bb  photometry.
The derived specific star-formation rates  for the bright LBGs are  $\sim$ 13 ~Gyr$^{-1}$, slightly higher  than  the 
 lower-luminosity LBGs, implying that the star-formation rate increases  with stellar mass for these galaxies.

  \end{abstract}


\section{Introduction} 

Discovering high-redshift  ($z>6$) galaxies, and  understanding their physical nature is  of great importance in studying 
the early  universe because these objects are likely sources for reionization of the  intergalactic medium (IGM:  
\citealp[e.g.][]{tre10, sal11,fin12})   at $z\gtrsim$ 7    \citep{fan02,kom11}.
  With the advent of large near-infrared (NIR) detectors,  significant  progress has been made in discovering galaxies at  
  $z \gtrsim7$  \citep[e.g.][]{iye06,van11,ono12,sch12},  as the rest-frame UV light from these galaxies shifts to $\lambda>$1~$
  \mu$m.

There are two main techniques for finding high-redshift   ($z\gtrsim7$) galaxies :  (1) the  Lyman-break  or dropout 
method  \citep[e.g.][]{ste95,ste99,ade04,dic04,gia04,ouc04,bou07,mcl10},   selected based on  strong absorption 
in their  spectral energy distribution (SED) at wavelengths  shortward of redshifted \lya  due to    IGM  HI  absorption,  and  
(2) the narrow-band imaging technique, which identifies  galaxies with a strong, redshifted \lya emission line  using 
narrow-band  filters    \citep[e.g.][]{mal02,rho05,iye06,kas06,ouc09b,hib10,til10,kru12}.
A unique advantage of  narrow-band selected galaxies is that these galaxies are more likely to  be confirmed via
spectroscopic followup because  they are pre-selected based on the  strong\lya  emission  line. On the other hand, LBGs 
 selected via the   \bb  dropout method are  unbiased in their selection (unlike  \lya  selected galaxies), and 
 LBG surveys  probe larger cosmological volumes due  to broader filters.  
 Currently there are many  candidate LBGs  \citep[e.g.][]{bou07,ouc09,mcl10}    at redshifts 
 much   greater than \zs\    \citep[e.g.][]{bou11,bra12,yan11,fin12,oes12}.

With growing samples of \zs\ galaxies, we are able to calculate the rest-frame UV luminosity function (LF), which is a simple yet 
powerful probe of the early universe. 
The luminosity  function  allows us to constrain the observed number  density of star-forming galaxies  from  which we can  estimate 
(with some assumptions) the number  of hydrogen-ionizing photons available to reionize the IGM.  
In addition, the observed luminosity  function allows us to study the galaxy evolution by  comparing the  number density of galaxies at 
different redshifts. 

Several studies have focused on understanding the evolution of the UV  luminosity  function   from $z\sim3$ to  6 and find that there is  a 
significant decrease  in the number density of UV bright galaxies  from low to high redshifts  \citep[e.g.][]{sta03,shi05,bou06}.
%  but  there are still large uncertainties on this value.  
If this trend continues, we expect even a greater  decline in the number density of galaxies at redshift $z>6$.
%Current observations  at $z>6$ suggest a strong evolution in the characteristic 
%magnitude $\rm  M^{\star}$;   at which the UV luminosity  function exhibits a  rapid change in faint-end slope  with  $\alpha< -1.6$ 
 %\citep{bou08,mcl10,ouc09,yan10,cas10,bou10a}.
Current observations  at $z>6$ suggest a strong   evolution in the characteristic magnitude  $\rm  M^{\star} \sim$ -19.8  mag
\citep{bou08,mcl10,ouc09,yan10,cas10,bou10a}  compared with $\rm  M^{\star} \sim$-20.9 mag at $z\sim4$  \citep{bou07}.
 Most of these studies, especially at \zs,   rely on \hst\   observations which have the advantage of probing fainter 
 galaxies,  however with smaller survey areas.  
 Recently,   \citet{ouc09}, \citet{cas10},  \citet{cap11}, and \citet{bow12}   found several  candidates using ground-based observations, 
  which have the advantage of  surveying larger areas and thus probing brighter, rarer galaxies.

 \begin{figure}[t!]
\vspace{-3.5cm}
\epsscale{1.25}
 \plotone{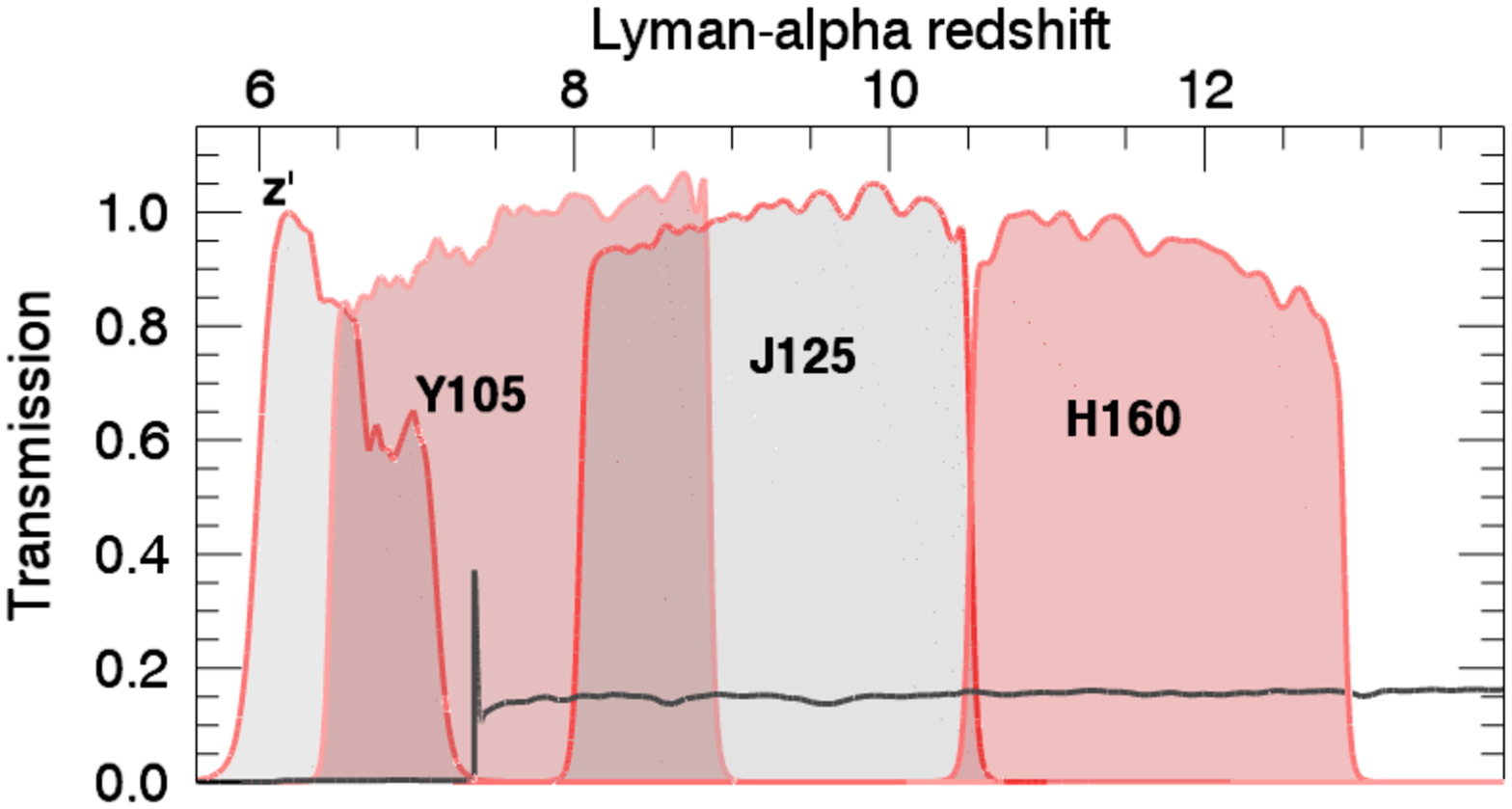}
 \vspace{-7.42cm}
 \epsscale{1.25}
\plotone{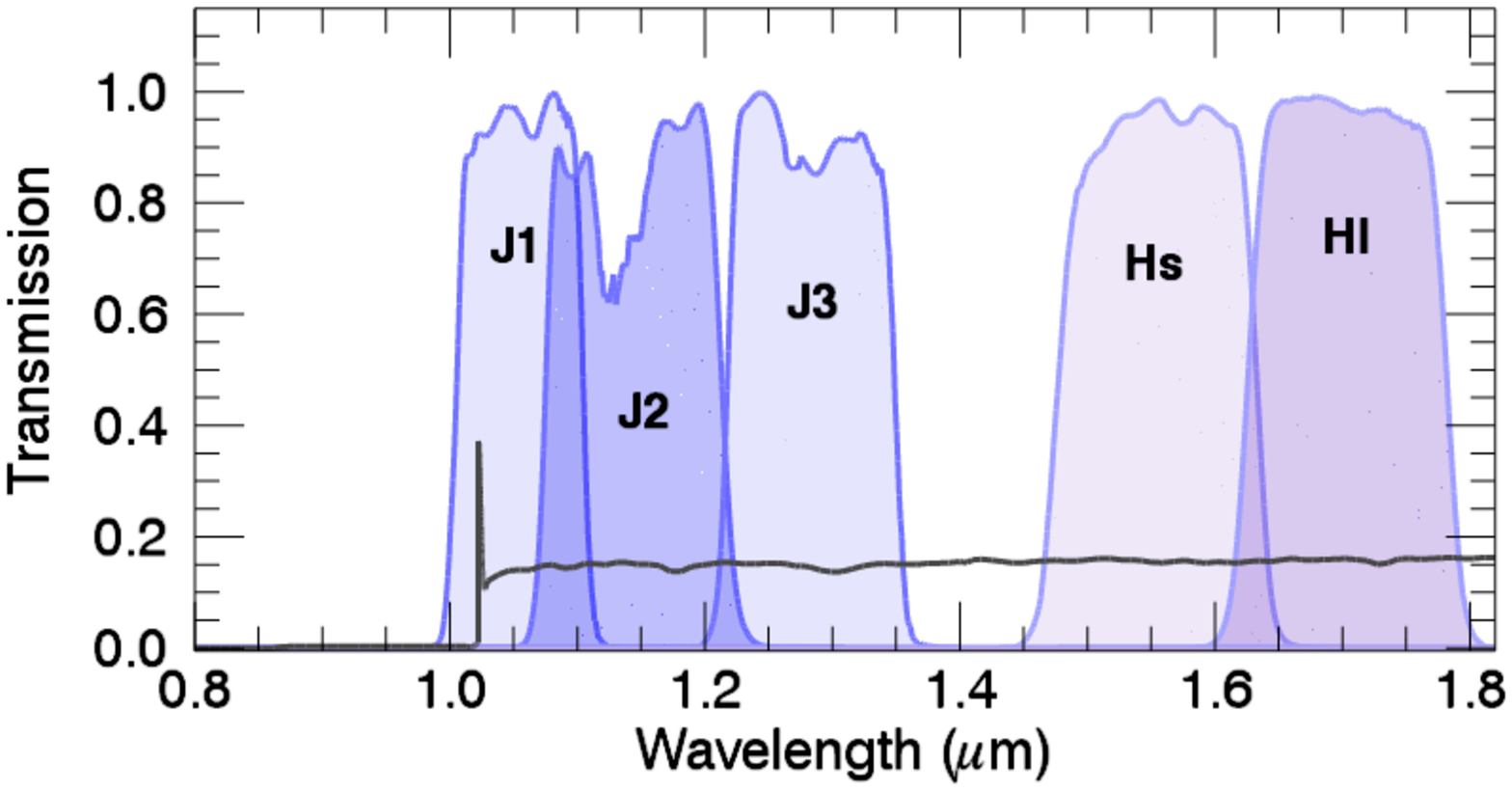}
%\epsscale{1.2}
%\vspace{-3.8cm}
%\plotone{filter_plot.eps}
%\plotone{bb_filt.eps}
\vspace{-3.95cm}

%\plotone{test.png}
%\rule{-14.5cm}{-0.2cm}

 \caption{Filter transmission curves of the \FS  \mb   compared to other \bb filters.  The top panel shows  
\bb Subaru  \zp,  \hst\ \yhst,  \jhst\   \& \hhst\   filters, while the bottom panel shows  \FS \mb  filters 
 \jone, \jtwo, \jthree, \hs\  and \hl, which provide R$\sim$10 resolution spectroscopy.
Black line shows a template  spectral energy distribution    of a $z=7.4$ galaxy with the \lya\  emission line falling in the \jone\  filter.   
%While the  central wavelength of  \jone\    and \hst\ $Y_{105}$ filters are nearly  the  same, the width of the \jone\   filter 
%is much smaller.   
%These  smaller widths of MB filters  are expected to yield narrower range of  photometric redshifts, compared to the 
%broadband derived redshifts.
The top axis marks the \lya\ redshifts as a function of wavelength.
 \vspace{7pt}}
\end{figure}

While it is relatively easy to construct the luminosity function, its accuracy  depends on a number of systematics, 
including the degree of contamination of the  \zs\  galaxy  sample. 
%systematics related to the purity of the \zs\  galaxy  sample.   
It   is affected by contaminants that mimic   the colors of high-redshift galaxies. 
The main  contaminants of \zs\ LBGs are Galactic brown dwarfs, low-redshift   ($z\sim 1-2$)  dust obscured  galaxies with very 
faint continuum in the visible   bands, and transient objects.   
The  brown dwarf stars may be  separated from (resolved) galaxies using high angular-resolution imaging.   
However, the ability to resolve galaxies   can become  unreliable at the  faint magnitudes and compact sizes typical of \zs\ galaxies, even 
with the  \hst\ observations.  For example, the rejection of point sources from  \zs\ samples precludes the possibility that 
some \zs\ galaxies have  morphologies that are unresolved  even at \hst\ resolution (FWHM  $\approx$0\farcs2  which 
corresponds to 1 kpc at $z=7$), leading to   sample incompleteness.  These uncertainties need to be accounted
for when constructing the luminosity function.

In addition to studying the UV luminosity function, understanding the physical  properties of  high-redshift galaxies is important to link 
their  evolution during the early stages of galaxy formation to galaxies at  other epochs.    
As we probe redshifts $z \gtrsim 2$, galaxies  on average have smaller stellar masses $\sim 10^{10}$ M$_{\odot}$
 \citep[e.g.][]{saw98,pap01,sha01,sha05,yan05,erb06a,fon06,red06,ove09}, bluer UV  colors 
  \citep[e.g.][]{dic03,pap04,lab07,ove09,bou09a},   and younger stellar populations with  smaller amounts of dust 
 \citep[e.g.][]{meu99,pap01,sha01,sha05,red05,red06,red08,red09,bou09a,fin10,fin11}.
%Moreover, galaxies at  $z \gtrsim 2$ appear to have   specific star-formation rates (sSFR: the SFR per unit stellar mass) that are 
%constant with
 %redshift:  galaxies with   $M > 2 \times 10^{9}$~\msol\ show sSFRs $\approx 2$ Gyr$^{-1}$ \citep[e.g.][]{red12}.   
 Moreover, the specific star-formation rates (sSFR: the SFR per unit    stellar mass) of galaxies at $z \gtrsim 2$ are roughly 
 constant as a function of redshift, with sSFR ranging from  $\sim~$2 to 6 Gyr$^{-1}$,
 %with typical values of $\sim$2  Gyr$^{-1}$, 
 albeit with large uncertainties 
 \citep[e.g.][]{sta09, gon10, bou11,  red12}.
 In all cases, the physical properties of the most distant galaxies ($z > 6$),  are derived by fitting  the spectral energy distribution
  templates  to the galaxy photometry measured in \bb filters
 \citep[e.g.][]{fin11}.
 While some  effects such as the degeneracy among different derived parameters are already known, the template fitting to the 
 \bb photometry   may introduce (as of yet)  unknown systematic errors on the constraints of  these parameters.

\begin{table}
  \centering
  \caption{Photometry   for COSMOS field.}
  \begin{threeparttable}
  \small{ 
      \begin{tabular}{llccc}
         \hline
         
	 Filter	&	$\lambda	(\mu m)$	& 	Depth (mag)\footnotemark[1]		&	Filter width \\
	 		&					&								& $(\rm \AA)$\footnotemark[3] \\
%	& ($10^{-17}  erg \; cm^{-2} \; s^{-1}$)	& time (hr)			 \\
	\hline
	\gp						& 	0.44 				& 	27.96	& 1384	&\\
	\rp						& 	0.62 				& 	27.95	& 1460	&\\
	$i$						& 	0.83 		  		& 	26.59	& 2270	&\\
	\zp						& 	0.90 				& 	26.43	& 1384	&\\
 	\jone						& 	1.05 				&	26.10 	& 1030	&\\
	\jtwo						& 	1.14   			& 	26.07	&1410	&\\
	\jthree					& 	1.28  			& 	25.69	& 1320	&\\
	\jhst{\footnotesize(F125W)}	& 	1.25  			& 	26.93	& 2991	&\\
	\hs						& 	1.55  			& 	25.15	& 1600	&\\
	\hl						& 	1.70  			& 	25.17	& 1610	&\\
	\hhst{\footnotesize(F160W)}	& 	1.54 				& 	27.04	& 2880	&\\
	\ks						& 	2.15  			& 	25.13	& 3380	&\\
	$[3.6]$\footnotemark[2]		&	3.56				&	26.0		& 7274\footnotemark[4]	&\\
	$[4.5]$\footnotemark[2]		&	4.52				&	26.0		& 9914\footnotemark[4]	&\\

\hline

 %subaru filter information from Taniguchi et al paper
      
        \end{tabular}}
     \begin{tablenotes}
     {\footnotesize 
       \item[a]  $5\sigma$ depth in 1\farcs1 diameter aperture
       \item[b]  $3\sigma$ depth in 2\farcs4 diameter aperture (Ashby et al.\ 2013 submitted)
        \item[c] Filter width where filter transmission is $>$50\%.
         \item[d] Transmission is   $>$30\%.
                \\}
     \end{tablenotes}
  \end{threeparttable}
\end{table}

\begin{figure*}
\epsscale{1.02}
\plotone{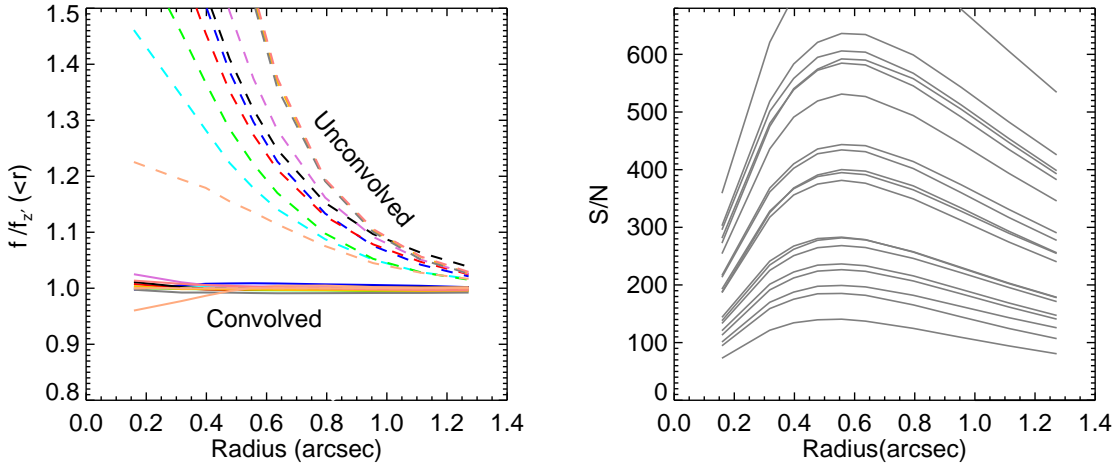}
 \caption{Left panel: 
 Curves-of-growth for point sources in the convolved and unconvolved images.
 Dashed  and solid lines show flux ratio before and after convolution, respectively, where 
 $f$  is the flux in a given band and  $f_{z^\prime}$ is the \zp\ (reference image)  flux enclosed within radius  r.
 Different colors indicate different filters.
The flux in  aperture radii larger than $0\farcs48$  among  convolved images  contribute
to $<2$\% uncertainty for point sources.
  Right panel:  The   signal-to-noise (S/N) for point sources as a function of radii, in the convolved images. 
Different lines show the S/N of same stars (here shown only the  \jthree\  filter for illustration purposes) that  were used for 
creating  the convolution kernel (Section 2).
For photometry,  we use a  circular aperture with  $0\farcs55$ radius   as this  yields  maximum S/N.\\
%Apertures with 3.5 pixel radii ($0\farcs55$) yield maximum S/N 
 %The figure shows that the S/N is maximized for point sources  with apertures of radii  3.5 pixels (radii $= 0\farcs55$).  
 %Therefore   we use circular apertures with diameter =$1\farcs1$ for the photometry in our analysis. 
   }
\end{figure*}

The specific goals of this paper are (1) to use near-IR  \mb  filters, for
the first time, to reliably  identify \zs\ galaxies  and to compare the
effectiveness of the \mb  technique to that of \bb photometry;
(2) to discuss the contamination rate of brown dwarfs in  \zs\ galaxy
samples;  and (3) to derive the physical properties of the galaxies
from the \mb photometry and to test for systematic uncertainties on
physical properties using \bb  data alone.   
Photometric redshifts derived using \mb  filters have much
narrower probability distributions compared to the \bb  derived
photometric redshifts owing to the higher spectral resolution (R$\sim$10) of the
\mb  filters.   This led to a recent discovery of  
a   distant ($z=2.1$) galaxy cluster in the  COSMOS field 
\citep{spi12}.
 Owing to  very deep imaging and excellent seeing, 
we are also able to place better constraints on the
redshifts and on the evolution of bright star-forming galaxies out to
\zs.  

This paper is organized as follows.  Section 2  describes
the observations, data reduction, and catalog generation.
In Section 3 we present our LBG color selection method and sources of
contamination and their identification.  
Section 4 describes the construction of the UV luminosity function and its comparison
to previous studies.
We
compare the advantage of using  \mb   over only \bb 
photometry, and its influence on the derived physical properties of
high-redshift galaxies in Section 5.  We summarize our conclusions in
Section 6.  Throughout this paper we use AB magnitudes and standard
cosmology with $h=0.7$, $\Omega_{\lambda}=0.7$,  and $\Omega_{m}=0.3$.

\section{Observations and Data Reduction}
We observed  a 155 arcmin$^2$ area in  the CANDELS\footnote{http://candels.ucolick.org/}/COSMOS field \citep{sco07}  
centered at RA: 10:00:31 
and Dec:+02:17:21, using the \FS  instrument, as part of the ongoing \FS Galaxy Evolution survey (\zf )\footnote{
http://z-fourge.strw.leidenuniv.nl/}      during 2011-2012.  
The \FS  instrument  (Persson  et al.\  2013) is a \nir  camera mounted on the Magellan/Baade telescope.  
It has 4096 $\times$ 4096 pixels with a   0\farcs159/pixel resolution   and covers a 10\arcmin.8$\times$10\arcmin.8 field of view.
We observed with five adjacent \mb  filters (\jone, \jtwo, \jthree, \hs,  \hl : Table 1) with a resolution of $R\sim 10$, and one  \bb  (\ks) filter 
(Labb{\'e} et al in preparation).
%
%to a $5\sigma$ depth of 26.10, 26.07, 25.59, 25.15, 25.17, and  25.14 mag respectively for a point source detection in $1.5"$ diameter aperture ( Labbe et al in preparation). 
%
Figure 1 shows filter transmission curves for the \FS  \mb  filters
compared to the \bb  filters from Subaru/Suprime (\zp), and \hst\ (\yhst,
\jhst,  \hhst) filters.
The \jone\ filter has a similar central wavelength to the \yhst\ filter (and also to many $Y$-filters on ground-based telescopes), but the 
\jone\ filter is much narrower compared with the \hst\   \yhst\ filter.

We processed all raw images  using a custom designed IDL-based
pipeline (described in more detail in  Labb{\'e} et al in
preparation) adapted from  the NEWFIRM Medium Band Survey data reduction pipeline  (NMBS:
\citealp{whi11}).
 The data quality of the \FS images is excellent and the point-spread function (PSF) FWHM  for the stacked  \FS
images correspond to 0.55, 0.52, 0.51, 0.51, 0.57, \& 0.48 arcsec in the \jone\ through \ks\ filters, respectively.
In addition to this data set, we used publicly available
data in the optical (Subaru \gp, \rp,  \zp, \& ACS $i_{814}(i)$), deep \nir\   \hst\
WFC3 \jhst\  \& \hhst\ from the CANDELS survey 
\citep{koe11,gro11},
 and infrared IRAC  
 \citep{faz04} data.
 We  also used  deep 3.6 and  4.5$\mu$m images (Ashby et al 2013 submitted) from the Spitzer Extended Deep Survey 
 ($SEDS$)\footnote{http://www.cfa.harvard.edu/SEDS/index.html};   5.8  and  8.0$\mu$m data\footnote{ http://irsa.ipac.caltech.edu/data/COSMOS/}
  from the  Spitzer 
 Space Telescope  \citep{wer04}.
 Table~1 presents the central wavelengths  and limiting magnitudes for the \zf  data as well as information for the ancillary data.

The PSF   among different bands varies
due to differing central wavelengths,  different
instruments  used for  imaging, and in the case of ground-based observations, due to different seeing conditions.  
We convolved all images to a
reference image having the broadest PSF-FWHM (in our case the Subaru
$z'$ image with PSF FWHM=0\farcs8).    In each band, we first constructed  an empirical PSF by identifying  stars and
stacking  them with their centers aligned.  We then used an IDL routine,
DECONV$\_$TOOL 
 \citep{var93},
 to construct a kernel 
to convolve the image PSF to match the stellar  PSF.     Figure~2 
%\todo{Casey moved this figure to here since there is nothing else in the appendix} 
shows the curve-of-growth of point sources in the images following the convolution process.  As shown in the figure 
(left panel), the relative flux for point sources in each band is matched to better than 2\% for all apertures with radii larger than 
$0\farcs47$.

The  photometry in circular apertures was optimized for
faint, compact sources.   Figure~2 (right panel) shows the S/N for point sources
measured in circular apertures as a function of aperture radius.   
The S/N is maximized for point sources with
apertures of radii  $ 0\farcs55$.  Therefore we
use circular apertures with diameter of $1\farcs1$ for the photometry in
our analysis.

\subsection{\spitzer/IRAC  Photometry}

As mentioned earlier, to extend the wavelength coverage of our candidates  and to derive the
physical properties as accurately as possible,  we   used  recent deep Spitzer
IRAC data (Ashby et al 2013 submitted)  in two bands (3.6 $\mu m$, 4.5 $\mu m$) 
obtained from the  SEDS survey. However, due to its low resolution, neighboring  objects tend to blend resulting in
incorrect flux measurements.  To circumvent this issue, we used GALFIT
 \citep{pen02}
code to measure the object fluxes, as described below.
% as accurately as possible, by subtracting nearby objects,
%and measuring the flux of the object of interest. 
%
%subtract the neighbouring galaxy, and then measure the flux
%in the residual image in both 3.5 $\mu m$ and 4.5 $\mu m$.
%
%\todo{Did we actually use any of the TFIT photometry ?  IF not, just
%cut this paragraph.}  We follow a method similar to the TFIT algorithm
%(Laidler et al.), and used by Labbe et al (2010) [check]. 
 First,  we fit  the two dimensional surface brightness profiles of 
 all objects within 7\arcsec\ of the \zs\ candidate  (excluding  the \zs\ galaxy) and 
 subtracted these fits to obtain a residual image.
   We then measured photometry of the \zs\ candidates 
    from  the IRAC  data using the IDL procedure APER,  measuring the flux in  a  fixed  2\arcsec\ 
 diameter aperture and correcting to the  total flux using the aperture corrections derived by taking the ratio
 between 2\arcsec\ and 15\arcsec\  diameter aperture fluxes of isolated sources.
 %  from the  Spitzer Science Center\footnote{
%http://irsa.ipac.caltech.edu/data/SPITZER/docs/irac/--}.
 Finally, we derived the uncertainties by measuring the $rms$ in 
 %
 %of the
%flux in circular annuli around the object.
apertures of the same size as above, randomly placed
in regions  devoid of objects in the IRAC images.

\begin{figure*}[t!]

\epsscale{0.95}  % may need to adjust this
\plotone{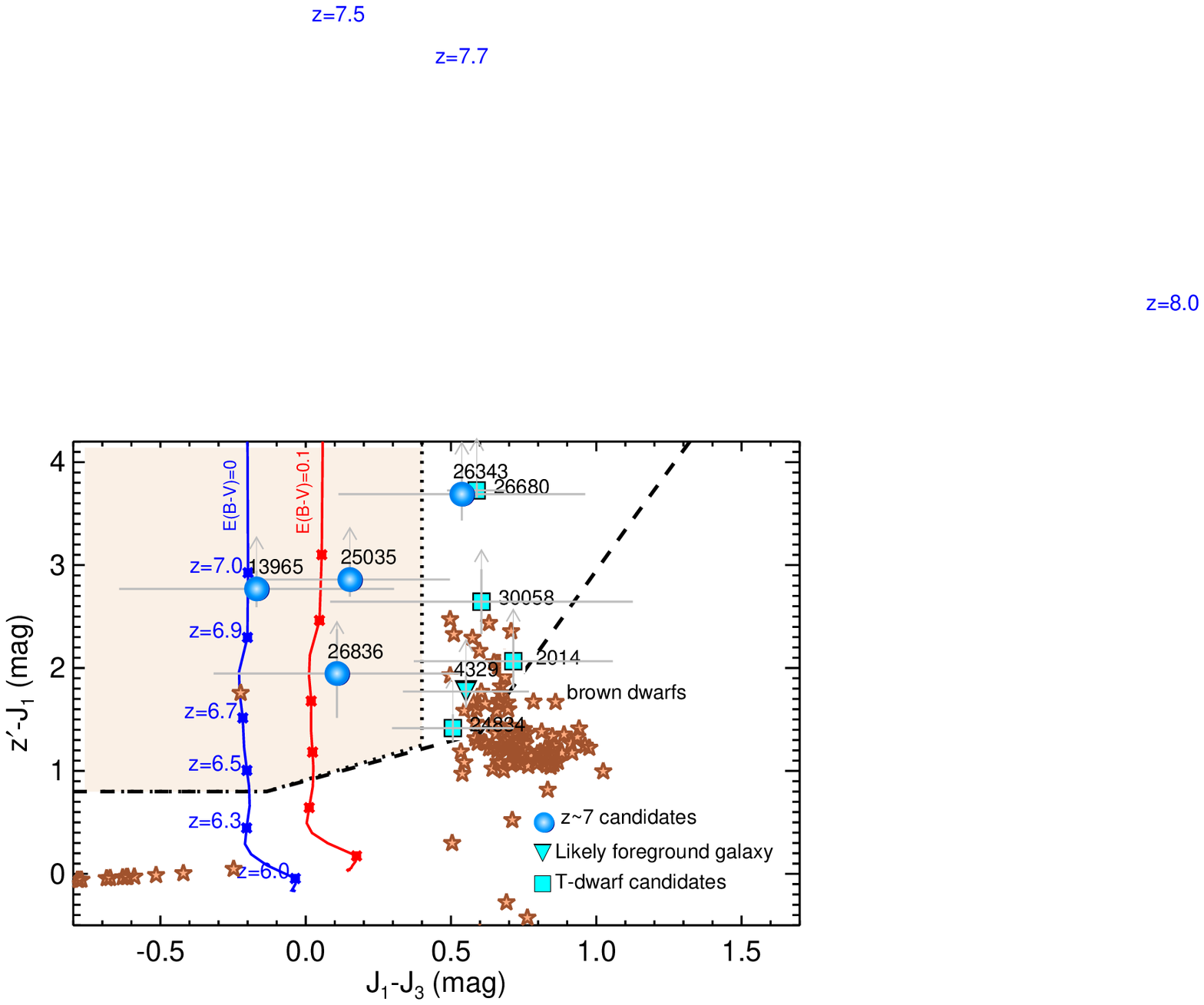}
 \caption{ Color-color plot for  \zs\  LBG candidate selection.
 Filled circles, triangle and  square symbols indicate all the objects that
pass the LBG selection criteria (Equation 1; dashed line), adapted from  Oesch et al.\ (2010).
Blue and red lines show the expected colors of a
star-forming stellar population, for two extinction values,  as a function of redshift (redshifts indicated for reference) using models from Bruzual \& Charlot
(2003).    
Star symbols show the expected colors of very late-type stars derived from  observed spectra of M, L, and \td stars.  
The initial color selection (dashed line) yields four  \zs\  LBG candidates (filled circles),  four brown dwarf stars (filled squares)
and one likely foreground galaxy (filled triangle).
Based on the galaxy tracks (blue and red lines ) and brown dwarf star colors, we refined the  $z\sim 7$ LBG selection  criteria 
(Equation 4; shaded region) that eliminates most of the nearby stars. This new selection yields three robust \zs\  LBG candidates.\\
 }
\end{figure*}

\subsection{Source Catalogs}

We identified sources in each of the images using Source Extractor software (SExtractor :  \citealp{ber96})
 in dual image mode. 
 In this mode,   a detection image  is used to identify the pixels associated with each object, while the fluxes 
 are measured from a distinct photometry image.   For the detection image, we used a $\chi^2$ image, 
 which optimally sums multiple images (accounting for varying S/N) for source detection   \citep{sza99}. The 
 $\chi^2$ image ($J_{chq}$) 
 was constructed from the \jone, \jtwo, \jthree\  images using  SWarp  \citep{ber02}. We measured object 
 photometry from the PSF-matched images  in a fixed aperture size of 1.1\arcsec\ diameter.
 % as this aperture   maximizes the S/N for compact sources.   
 %For the sources of interest here, this aperture size is slightly 
 %greater (5 kpc per arcsec at $z$=7) than the   expected half-light radii of \zs\  galaxies.

While SExtractor estimates the noise for each object, this  estimate is incorrect  in the convolved images  as 
it does not account for  correlated noise, which results from  the image reduction and convolution processes. 
To estimate the flux uncertainty for each object  in the convolved images we followed  an empirical approach 
described in \citet{lab03}.
 To estimate the pixel-to-pixel noise we measured the rms background variation as a function of linear size  
 $\rm N= \sqrt{A}$ where $A$ is the area of an aperture.  First, we measured the sky flux in increasingly larger 
 apertures that are randomly placed in the PSF-matched-convovled image, taking care to avoid detected objects.
From the distribution of fluxes measured in a given aperture size, we measured the standard deviation, $\sigma$. 
The 1$\sigma$ error on the flux was  then calculated using  equation 5 from \citet{whi11}. Our final catalog 
contains aperture fluxes obtained from  PSF-matched images in  \gp, \rp,  $i$, \zp,  \jone, \jtwo, \jthree,
\jhst,  \hs, \hl, \hhst,  and  \ks\ along with flux uncertainties.

\subsection{Photometric Redshifts}
 We derive photometric redshifts using the photometric redshift code EAZY  \citep{bra08} taking advantage of extensive  
 multi-wavelength data   publicly available  in the COSMOS field
%
%The reliability of photometric redshifts however, depends greatly on the wavelength coverage as well as the number of SED templates used
%
for finding the best-fit solution to the observed data.
%
%As S/N is lower for fainter sources, the photometric redshift estimates tends to be unreliable for faint sources.
%
The EAZY code \citep[v2.1;][]{bra11} contains seven different templates spanning a large range of galaxy colors and  
incorporates nebular emission lines.  While our \zs\ candidate selection is based primarily on a color-color selection,
to increase the reliability of our candidates,
% in addition to the color-selection criteria,    
we also require that each of the candidates have  photometric redshifts $z>6$.

\section{LBG Color Selection  at $z\sim7$}

Our initial color-selection criteria to identify candidate galaxies at 
$z\sim7$ is  adapted from  
 \citet{oes10};
%\begin{equation}
\begin{align}
 \mathrm {S/N }(J_{chq}) &> 7.0 \\
 \mathrm {S/N (optical)} &< 2.0,\nonumber \\
  \jone -\jthree & < 1.3  \; \rm {mag},  \nonumber\\ 
  z' -\jone &  >0.8 \; \rm {mag},\nonumber\\
z'-  \jone &> 0.9 + 0.75(\jone - \jthree) \; \rm {mag},  \nonumber\\ 
z'- \jone &> -1.1 + 4(\jone - \jthree) \; \rm {mag}.\nonumber
  \end{align}
%\end{equation}
%$\rm S/N ($J$_{chq} > 7.0 $,\\
%$\rm S/N (optical) < 2.0$, \\
%$\rm \jone -\jthree < 1.3 \; mag $, and \\
%$z' \rm (mag)-\jone(mag) >0.8$.\\
%
Similar criteria  have been used in other studies to identify \zs\
galaxies 
 \citep[e.g.][]{bou09a,ouc09,cas10}.
In addition to these criteria, we required that each of the candidates be detected in the 
 \jtwo\ filter as this filter is very effective in discriminating \td  stars (Tilvi et al in preparation), due to their methane and
 water absorption features \citep{tin12}, 
 from high-redshift star-forming galaxies.

Figure~3 shows a \jone-\jthree\   vs  \zp-\jone\    color-color plot and illustrates the region (dashed line) that corresponds 
to the selection criteria of Equation 1.   Color-color tracks of star-forming galaxies  with two different dust attenuation are 
shown in blue and red lines with redshift labels.  These tracks are obtained from \citet{bru03}  stellar population synthesis
models by integrating the model spectral energy distributions with the \zp, \jone, and  \jthree\  filter.  The models assumed E(B-V)= 0.0 (blue line) 
and  0.1 (red line)  with Z=0.2Z$_{\odot}$ metallicity, constant star-formation history, IGM opacity taken from \citep{mei06}, 
and a constant age of $t$= 100 Myr. 
Based on these color tracks it can be seen that the star-forming galaxies occupy  a specific region of the color-color
space.

We identified 9 objects in total satisfying the selection criteria from Equation 1. Based on several tests  described in  
the following sections, out of the 9 objects in the initial sample (hereafter initial  source list),  four objects are identified as
\zs\ LBG candidates, four objects  are identified     as brown dwarf stars and the remaining object is identified as a  likely
foreground dust-obscured galaxy at $z < 3$.
To minimize the contamination from stars and foreground galaxies, we further refine the color-color selection (Section 3.3) 
that include only the robust \zs\ LBG candidates (Figure~3; shaded region).
Figure 4 shows postage stamps of these   \zs\ LBG candidates, as well as one  likely \td   star for comparison.

%After visual inspecting, we removed three 
%objects as these three objects where bright saturated stars in the optical bands.
%Figure 3 shows postage stamps of all objects that are selected as $z\sim 7$ dropout galaxies. 
%All the candidates are detected in at least one of the medium-band J filters, and undetected $(<2\sigma)$
%in the optical bands.

\subsection{Sources of Contamination}
The color-color selection of \zs\  LBG candidates suffers from
several possible sources of contamination including image artifacts (or spurious
detections), nearby cool brown dwarf stars which mimic high-redshift
galaxy colors,  intermediate redshift galaxies with faint continuum in
the visible  bands, and transient sources including supernovae
or asteroids.  In this and the following sections we discuss in detail
the likelihood of any of these contaminating sources entering the initial source list.

\subsubsection{Spurious detections} 

In the \zs\  LBG  selection,  we require that each of the objects is 
detected with S/N $> 4$ in more than one band redward of the \zp\  filter.  
This requirement makes spurious detections due to either background or electronic noise extremely unlikely.
\vspace{1cm}

\subsubsection{Transient/High-proper motion objects} 

It is possible that transient objects including asteroids, and
supernovae can be  detected in more than one band.  However, with 
the availability of  two  epoch data, separated by about one year, we can eliminate any
transients by comparing their positional shift between the two epochs.
Therefore, we conclude that the initial source list  does not
contain any transient sources.
%
 %  In fact, we found one transient object (FS6752), likely an asteroid, which was removed from our \zs\  candidate list after comparing the 
   %two epoch data. \vspace{0.2cm}

   \begin{figure*}[t!]
\epsscale{0.58}
   \plotone{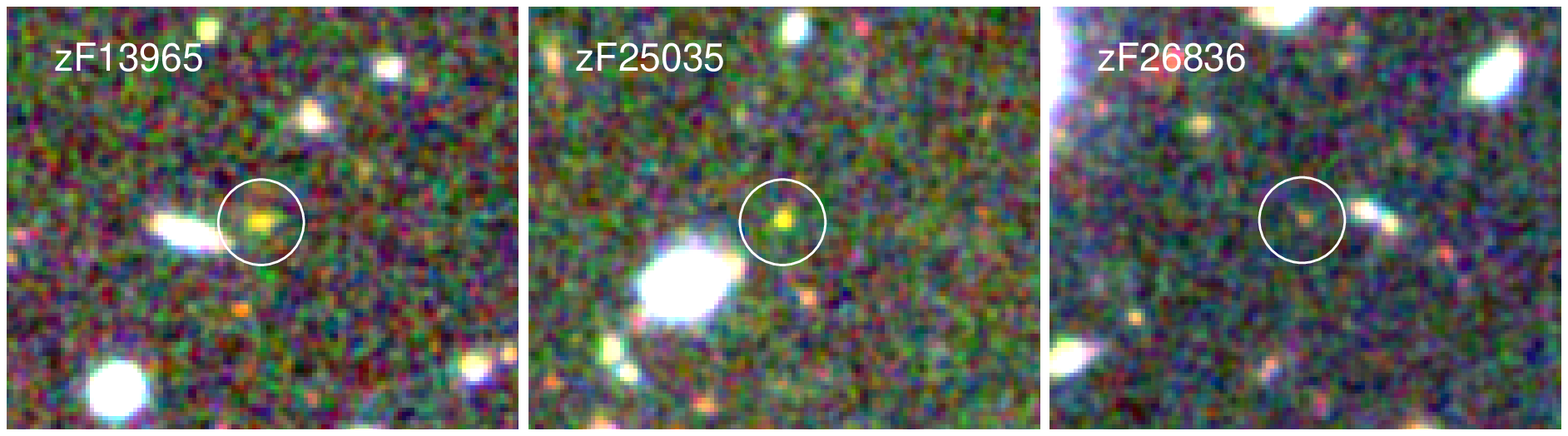} \hspace{0.4cm}
    \epsscale{0.2}
   \plotone{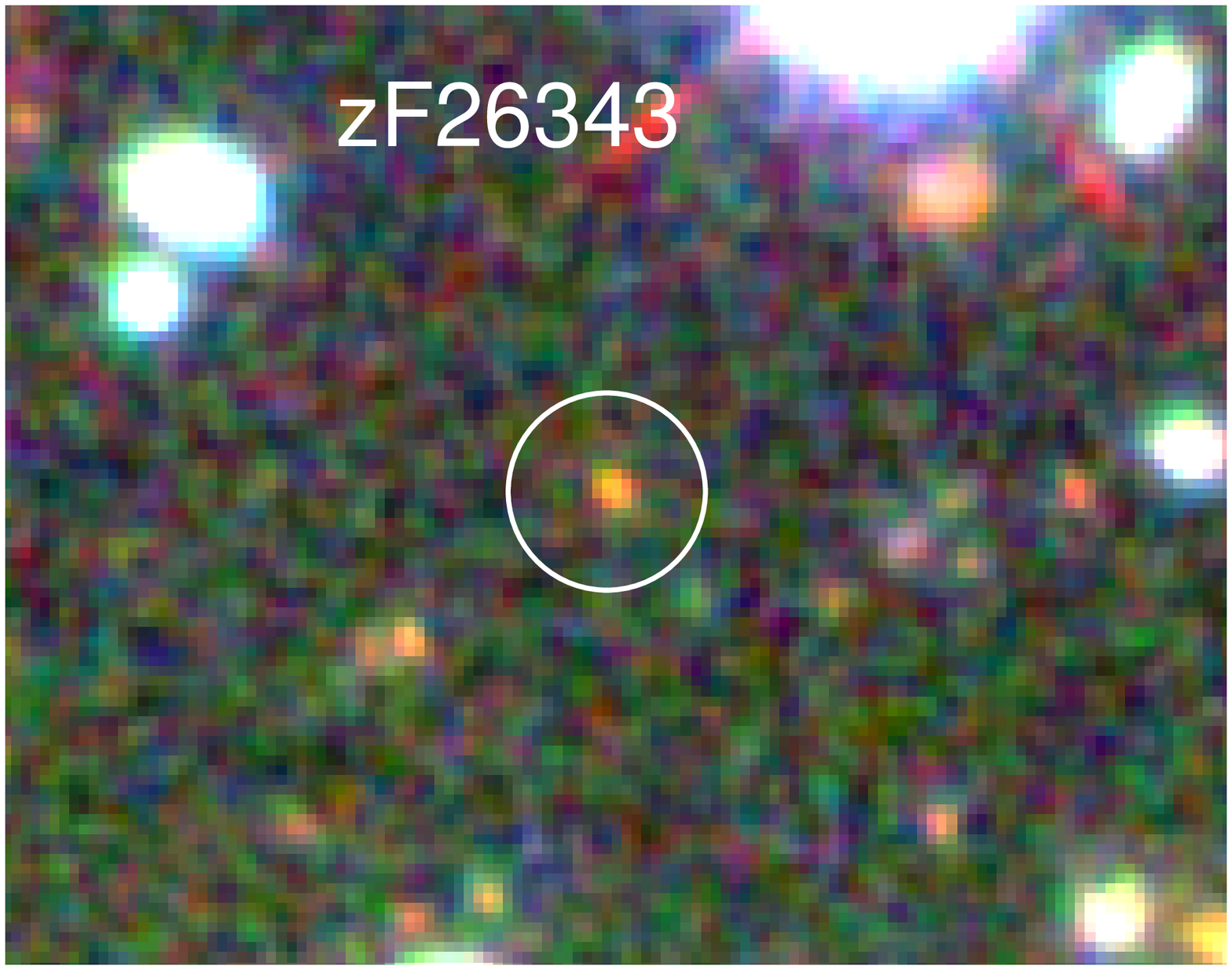} \hspace{0.4cm}
 \epsscale{0.2}
   \plotone{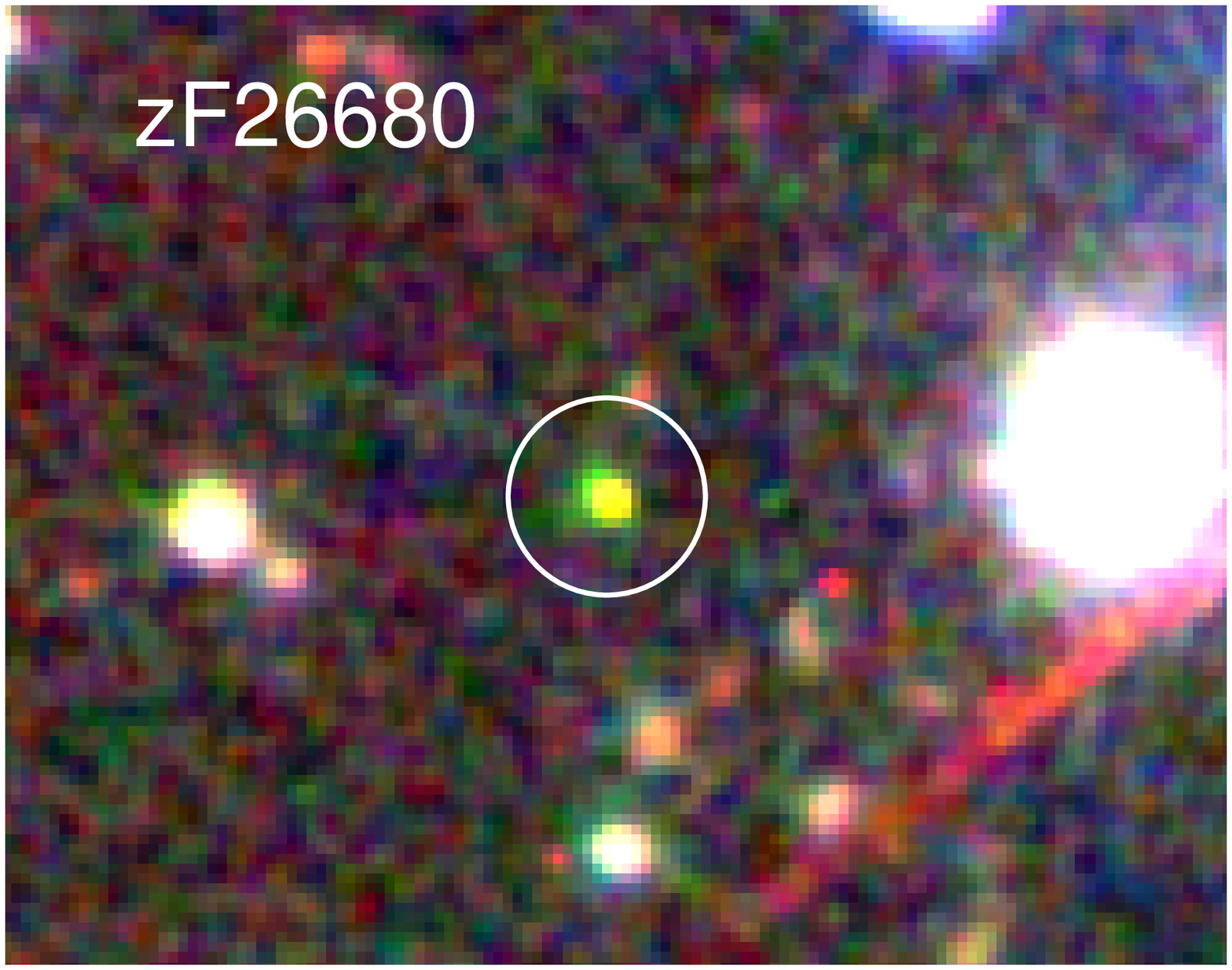} 
   
\epsscale{1.09}
\begin{verbatim}
         g'     r'     i      z'    J1     J2    J3    J125    Hs     Hl   H160     Ks    [3.6] [4.5]  [5.7]   [8]
\end{verbatim}
    \vspace{-0.5cm}
%\spaa \spaa \spaa \spaa \gp  \spaa   \rp\  \spaa  i  \spaa \zp\  \spaa J1   \spaa  J2  \spaa   J3  \spaa F125  \spaa Hs \spaa   Hl  \spaa  F160  \spaa Ks  \spaa ch1 \spaa 
 % ch2 \spaa  ch3 \spaa  ch4 \spaa \\
%\hbox \spread \linewidth{  \gp   \rp\  i \zp\ J1   J2   J3 F125 Hs  Hl  F160 Ks ch1 ch2 ch3 ch4 }  \\
%{\hspace{9mm} \gp \hspace{7mm}  \rp\hspace{6mm}   i\hspace{8mm}  \zp\hspace{6mm} J1\hspace{5mm}   J2\hspace{5mm}   J3\hspace{4mm}  F125\hspace{4mm}  Hs\hspace{4mm}  Hl\hspace{5mm}  F160\hspace{4mm} Ks\hspace{3mm} ch1\hspace{4mm} ch2\hspace{4mm} ch3\hspace{4mm} ch4}\\
\hspace{0.1cm}
  \begin{sideways} {\scriptsize   zF26836\spa  zF25035 \spa    zF13965 }\end{sideways}
\plotone{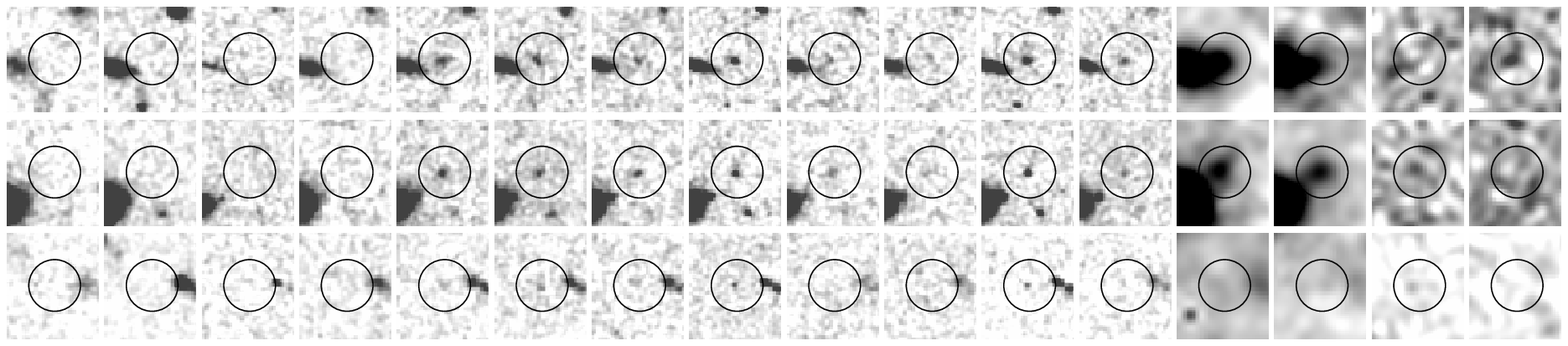}
 \begin{sideways} {\scriptsize   $\;$ \psix=0.99   \psix=0.99  $\;$    \psix=0.99    }\end{sideways}\\
\vspace{0.2cm}

\hspace{0.13cm}
\epsscale{1.093}
  \begin{sideways} {\scriptsize   $\;$zF26343 }\end{sideways}
     \plotone{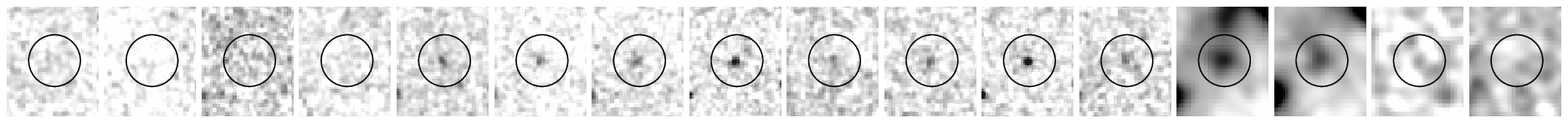}
   \begin{sideways} {\scriptsize   $\;$ \psix=0.99 }\end{sideways} 
  \vspace{0.2cm}
  
  \begin{sideways}    $\rm \;Tdwarf$  \end{sideways}
  \begin{sideways} {\scriptsize   $\;$zF26680 }\end{sideways}
  \epsscale{1.076}
 \plotone{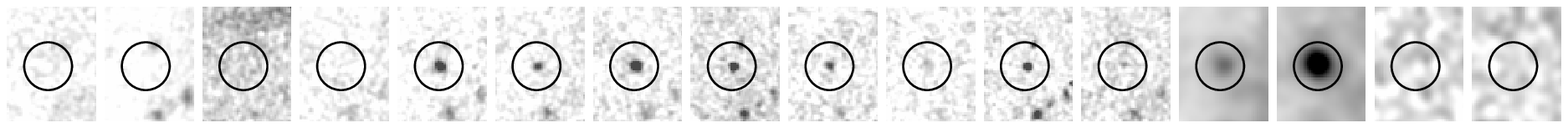}
  \begin{sideways} {\scriptsize    \psix=0.99 }\end{sideways}
  \hspace{0.2cm}
 \caption{Image cutouts of four \zs\  LBGs and one of the \td candidates (zF26680).
 Top panel shows color composite (\zp,\jone,\hhst) image cutouts while lower two panels show 
 image cutouts in the individual filters.
 The  four   $z\sim 7$ LBG candidates (zF13965, zF25025, zF26836, zF26343) ) are selected from the  color-color selection 
 (Equation 1) while zF26343 is excluded in the refined color-color selection (Equation 4).
 Each candidate is marked within  a 2\arcsec\  radius circle.
  These candidates are clearly detected in at least one of the \mb  filters while undetected ($<2 \sigma$) in any of the
  visible bands.  
The integrated probability distribution \psix\  (Equation 2) is shown on the right side. 
In addition to satisfying   the color-color selection criteria, each of the candidates is  required to have \psix $>0.7$ and 
S/N$<2$ in the visible $\chi^2 $ image.
These two criteria minimizes the contamination from low-redshift galaxies.
%The middle panel shows the \zs\  candidate that falls outside our refined color selection.
In the bottom panel, we show one of the  candidate \td   stars that was selected in the initial 
color-color selection (Oesch et al 2010). Such \td  stars can be cleanly distinguished from galaxies using  \mb 
 photometry.\\}
\end{figure*}

\subsubsection{Foreground galaxies}
While the  \zs\   candidates are required to have no detection ($<2 \sigma$) in the observed  visible bands, it is possible that
 the initial source list contains foreground  objects,  likely $z\sim1-3$ highly dust obscured galaxies that have very low  
 fluxes in the visible bands, but are bright in the \nir  and IRAC bands, (see Labb{\'e} et al 2005, Papovich et al 2006). These low-redshift 
 galaxies  cannot be completely ruled out, but the contamination from such objects can be minimized because they
 occupy a specific  color-color space (Figure 3) compared to the \zs\  galaxies which have bluer colors.  Furthermore, 
 many of the red, intermediate redshift galaxies at $z\sim 1-4$  have bright IRAC fluxes  with typical colors $[4.5] - [5.8] 
 \lesssim 0$~AB mag   \citep[e.g.][]{yan04}, which distinguishes  them strongly from \zs\  galaxy candidates.  
 In addition to these signatures,  the photometric redshifts can provide an additional constraint on the likelihood of an object  
 being a low-redshift galaxy.

In order to  compute the likelihood of our candidates being  \zs\  galaxies, we followed a similar procedure, as described in  
 \citet{fin10}.  We  compute the total redshift probability distribution $\mathcal P_6$ given by  
\begin{equation}
\mathcal P_6 = \frac{\int_6^{9} {\tiny P(z)\, dz}} {\int_0^{9} P(z)\, dz},
\end{equation}
where $\int_6^{9} P(z) dz$ is the integrated probability distribution from $z=6$ to $z=9$.  This probability distribution is 
normalized by $\int_0^{9} P(z) dz$ such that $\mathcal P_6 \le1$.  We make use of the  probability distribution  $P(z)$ 
derived using the photometric redshift code EAZY  and require that each of the  \zs\  candidates has $\mathcal P_6>0.7$. 
This means that each of the \zs\  candidates has greater than $70\%$ probability of being a   $z>6$ galaxy, based on the photometric
redshift. All of the 9 objects in the initial 
source list have $\mathcal P_6>0.99$ and 
%  Moreover,  all objects that we select using  Equation 1  have 
%$\mathcal P_6 > 0.99$.   
in principle,  we could restrict the sample to the stricter  requirement of  
 $\mathrm P_6 > 0.99$, but this  would have risked  missing some \zs\  galaxies near  our detection limit (see  \citealp{fin10}).

\subsubsection{Visible $\rm \chi^2$  Image}

To further  test the reliability of \zs\  candidates' non-detection in the individual visible bands, we combine \gp, \rp, and $i$ 
images to construct a visible $\rm \chi^2$  image; the combined $\rm \chi^2$ image is significantly deeper than individual
images.
We then run SExtractor to measure the S/N of all objects in the $\rm \chi^2$  image. Based on S/N($>2$)  in the visible
 $\rm \chi^2$  image, we reject one object (zF4329)  out of 9 objects from the initial source list,  as being a likely  foreground 
 galaxy   with very faint continuum.  This rejection is also supported by its strong detection in the MIPS \citep{rie04} 24$\mu m$ flux;
 there are no MIPS detections for  any of the remaining objects.

\subsection{Distinguishing  Stars from Compact Galaxies}

Galactic brown dwarfs ($M$, $L$,  and  $T$ ) are one of the main contaminants because  their \nir colors  resemble 
high-redshift (\zs) galaxy colors. While it is possible to identify candidates for stars using their FWHM and stellarity 
index (produced by   SExtractor) when using  \hst\ observations, these classifiers tend to  become unreliable at fainter 
magnitudes. In addition, at high-redshifts we expect  some galaxies to be very compact with their FWHM ($<1$ kpc at
\zs) comparable to the PSF FWHM.  In such cases, even with \hst\ observations, it would be  challenging to discriminate 
between stars and compact galaxies.

\begin{figure}[ht!]
\centering
\includegraphics[scale=0.8]{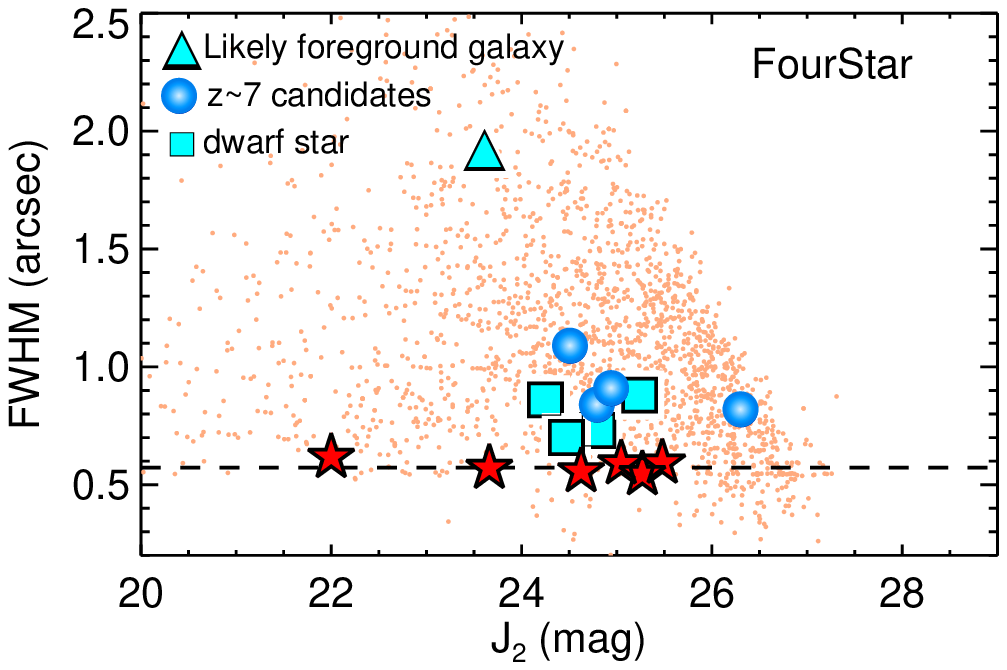}
\label{fig3.1}
\vspace{-0.75cm}
\centering
\includegraphics[scale=0.8]{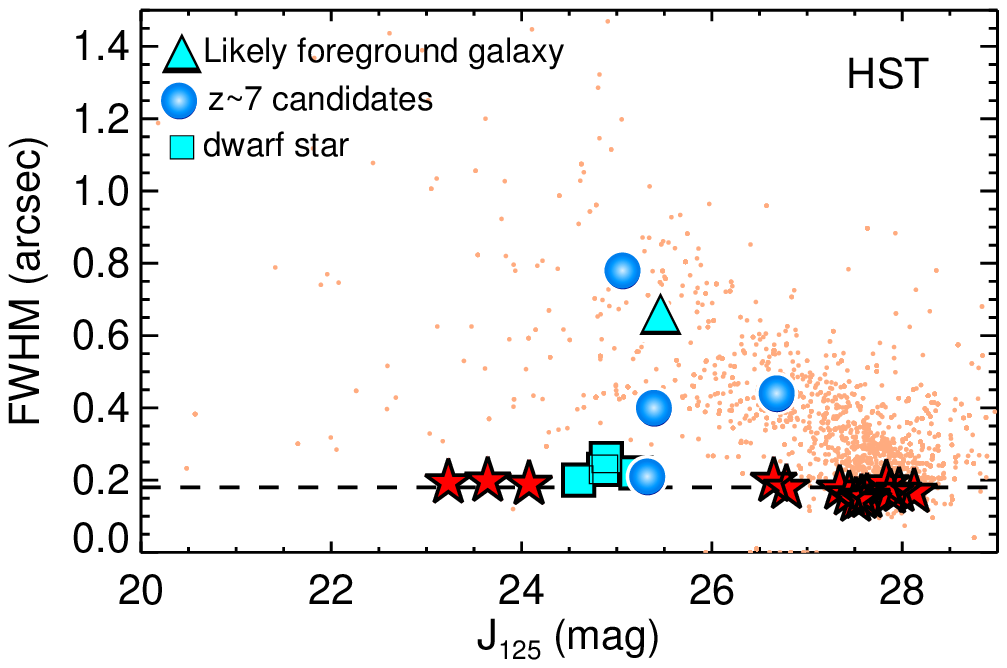}
%\caption{Chain Topology with Two Hops}
\label{fig3.2}
%\end{minipage}
%
\caption{FWHM  as a function of magnitude. Stars (star symbols) are identified based on their FWHM  and  stellarity index 
$ >$0.8 extracted from SExtractor.   Filled  symbols (same as in Figure~3) show all objects that passed the initial \zs\  color selection. 
Top panel: \jtwo\ mag vs FWHM  from the \FS bands (ground-based observations). 
Bottom panel: \jhst\  mag vs FWHM (space-based observations).   The distinction between stars and compact high-redshift 
galaxies, based on their FWHM,  is unreliable  from ground-based observations, compared with HST observations.
%As can be seen from the top panel, it is difficult to distinguish between compact high-redshift galaxies and
%stars simply based on FWHM from the ground-based observations.  On the
%other hand, FWHM from WFC3 bands are reliable in identifying stars, at
%least the bright ones.  
All fours dwarf stars (square symbols) have stellarity indices $>$ 0.7 while object zF25035 has a stellarity index of 0.6 
despite being its FWHM similar to the stellar FWHM. 
Small dots represent all other  objects  in the catalog. For clarity purposes, we show only a few representative  objects.
Note that the \jhst\ magnitude for the foreground object (triangle)  is significantly fainter compared with the \jtwo\ magnitude due to the fact that
in the raw (0\farcs06) \jhst\ image, this object is identified as three different objects.
\\}
\end{figure}

\begin{figure*}[t!]
\centering
\mbox{\subfigure{\includegraphics[width=3.35in]{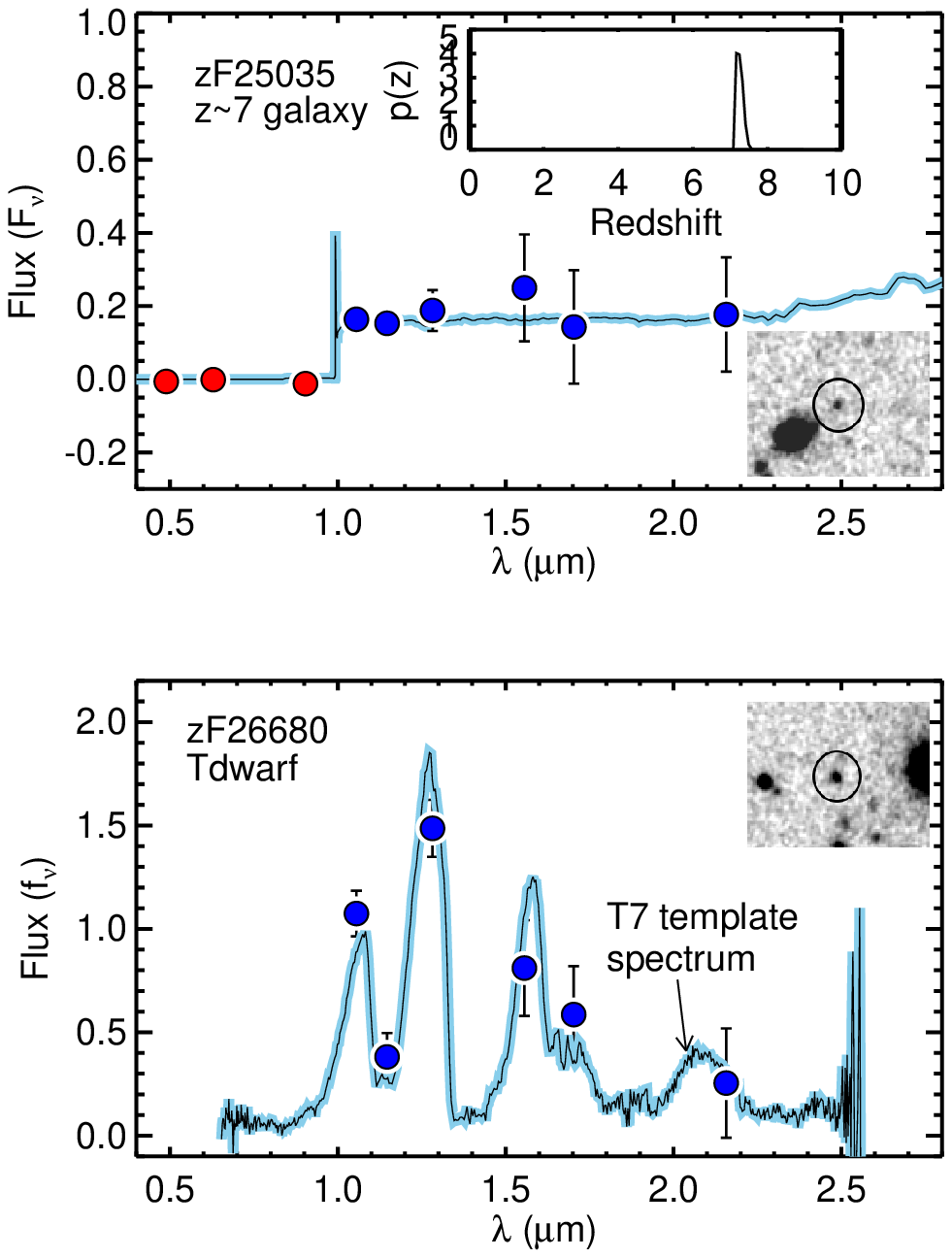}}\quad
\subfigure{\includegraphics[width=3.2in]{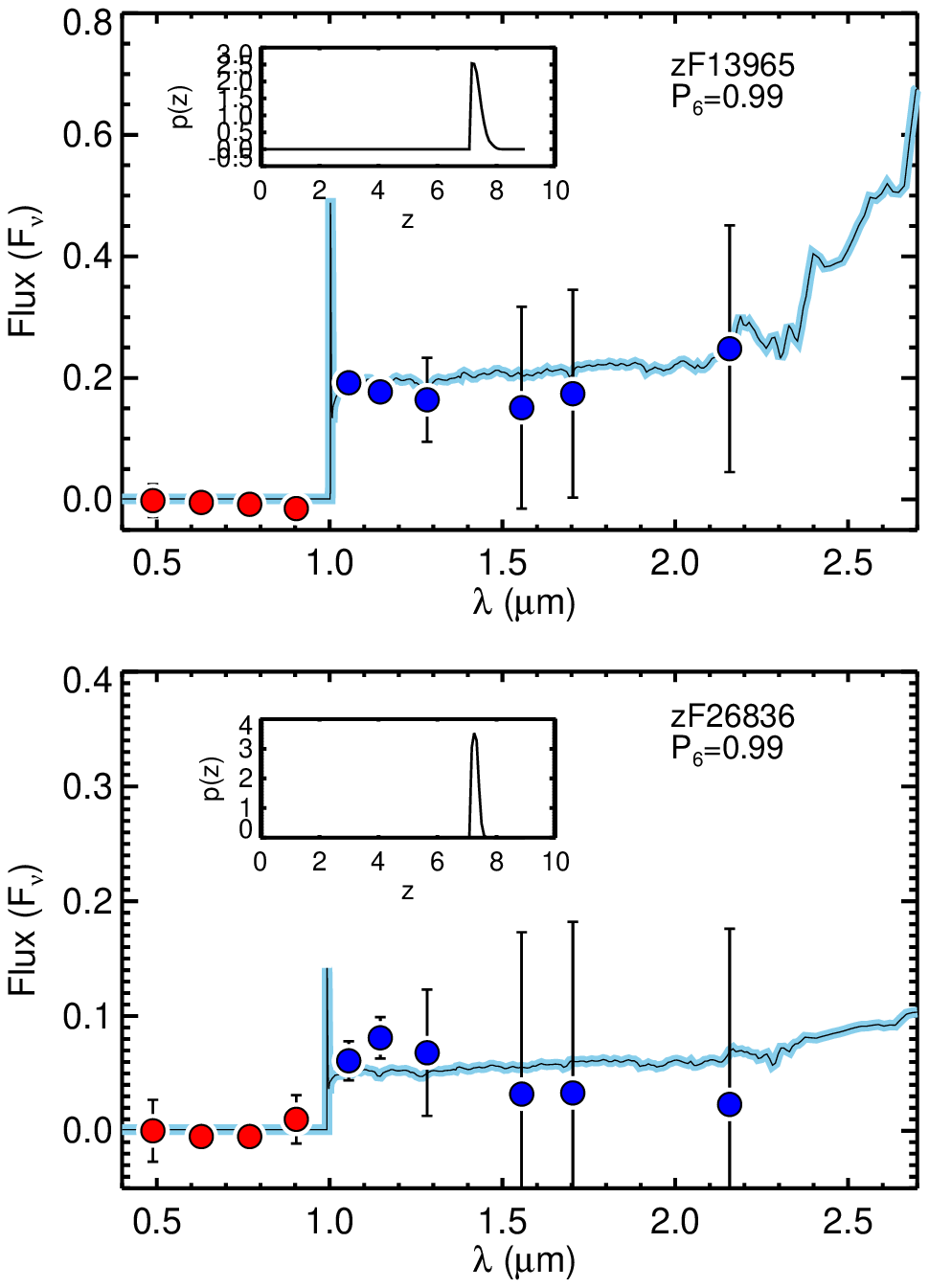} }}
\caption{Spectral energy distributions (SEDs)  of the three \zs\  LBG    candidates  and one  of the \td  candidates.  For all 
panels the filled circles represent observed photometry.  For the three panels showing the \zs\   LBG  candidates (zF25035, 
zF13965, zF26836) the solid line shows the best-fit SED template   and the inset panel shows the photometric probability 
distribution obtained from the EAZY code. In the left panels, also shown are the image cutouts of the LBG candidate (top left)
and T-dwarf candidate (lower left panel).
For one of the  the \td  candidates (lower left panel), the solid line represent the best-fit  
brown dwarf template spectrum, derived  by  minimizing   $\chi^2$ between the \FS photometry and the expected flux density 
synthesized from spectral templates of M, L,   and  \td stars from Burgasser et al. (2006).  
The \mb  photometry clearly traces the \td  observed template spectrum (lower left panel) and can  distinguish between 
different  \td types (Tilvi et al., in preparation). \\ }
\end{figure*}

In the following subsections, we perform three tests to distinguish stars from compact galaxies:  (i) tests comparing the objects' 
FWHM and stellarity indices, (ii) tests of the objects' spectral features indicative of molecular absorption in stellar atmospheres
of brown dwarf stars, 
and   (iii) tests of the objects' surface brightness profiles.

\subsubsection{Identifying Stars using FWHM and  Stellarity Index}

Figure 5  shows  the FWHM  (obtained from SExtractor) for objects as a function of \jtwo\  magnitude  as well as the FWHM
 measured in the (unconvolved) \hst\ \jhst\ band  to compare ground-based data with the space based observations. The 
 horizontal dashed line in each panel of  Figure~5 indicates the stellar PSF FWHM obtained by stacking  ten 
 isolated stars in the respective  images.
 
In addition to the FWHM, SExtractor produces a  stellarity index (CLASS STAR),  which determines a likelihood  of an object  
being either a point or an extended source   using the self-training of  a neural network.  A stellarity index closer to 1 is likely a point 
source while extended objects like galaxies have stellarity indices closer to 0. Figure~5 (top panel), shows  some objects 
classified as stars based on their high ($>0.8$) stellarity index (star symbol).   All objects in the initial source list  are shown in
filled circle, box, or triangle symbols.  As can be seen from the top panel, most of the objects appear to be resolved in the \FS
image.   However, this appears to be due to the  seeing conditions and the magnitude limit of  the ground-based data.  The lower panel in Figure~5 
shows that five of the objects in our sample (zF2014, zF24834, zF30058, zF26680, and zF25035) have FWHM in the HST 
image much closer to that of a point source.   Four of these objects (square symbols in Figure~5) have high stellarity 
indices ($>$0.7), indicating that they are likely point sources.   
In fact, fitting brown dwarf spectral templates to the 
\mb   observations (Section 3.2.2), we conclude that all four of these are candidates for brown dwarf stars.
The remaining  compact source, zF25035 has a stellarity index  of 0.6,  suggesting that it might be a compact galaxy rather than a point source. 
%
% Therefore, it is unclear from its FWHM and stellarity index  if zF25035 is a point %source or a compact galaxy.  
%
In the following sections, we discuss additional tests to distinguish between the possibility that this object is a point source, or a very compact \zs\  galaxy candidate.
 
 %But, as mentioned earlier, stellarity index tends to  be un-reliable at fainter magnitudes.
 %In the following sections, we do additional tests to confirm this.
% However, interestingly for object FS12030, we do not find any brown-dwarf spectra that fits its observed photometry.

 % are in fact nearby T-dwarf stars that can clearly be identified using the medium-band filters. The remaining one is likely a compact
  %\zs galaxy (section 3.3).
 %Thus, it is clear that using ground based observations, stars and galaxies can not be cleanly separated from each other.
 %On the other hand, bright stars can be 

\subsubsection{Spectral Fitting to Medium-band Photometry}

Due to the availability of  \mb   photometry, in addition to using stellarity  indices and   FWHM, we can use a spectral fitting 
technique whereby observed spectral templates are fitted to the \mb   photometry.  This method works very well especially in 
identifying \td  stars because the central wavelengths of the \FS \mb   filters trace the strong absorption features of \td stars;  
this  would not be possible using \bb photometry.   \citet{van09} have shown one of the ways to identify \td stars using \mb  
photometry from NEWFIRM (see their Figure 5).
%
%Tdwarf stars are among the coolest stars characterized by  $\rm H_{2}O$ and $\rm CH_4$ absorption in their 
%spectra . Due to these absorption features, they can mimic high-redshift galaxies that are selected 
%by Lyman Break method.
%But due to the availability of \FS medium-band filters, we can clearly distinguish between compact
%galaxies and stars.

To identify dwarf stars, we synthesized \mb  photometry from the M,  L, and \td  observed spectra   \citep{bur06}
and fit these to the \FS  photometry of all  candidates to obtain the minimum $\chi^2$ between the candidate  photometry and the 
dwarf template spectra. Figure 6 demonstrates the  effectiveness of \mb   filters in identifying \td stars and best-fit 
spectral energy distribution (SED) of \zs\  
LBGs.   The bottom-left  panel shows the best-fit spectrum of an observed \td  star, which clearly traces the \mb   photometry,
while the remaining panels show  \zs\   candidates fitted with  galaxy SEDs. Using this method,  we classify  four objects, 
previously identified as stars based on their \jhst\ FWHM,  as  dwarf stars.  
For the  compact   object (zF25035) however,  we did not find a good fit with any of the M, L, or  \td observed spectra. 
Thus, this test favors the conclusion that this object is a compact \zs\  galaxy.

\subsubsection{Surface Brightness Profile}

Figure 7 shows the surface brightness (\jhst) profiles of five  objects: two \td  candidates (with \mb colors indicative 
of brown dwarfs and  high stellarity);  two  \zs\  candidate galaxies including the compact  zF25035,  and a  random object
 selected from  the subsample of all objects with stellarity $>0.9$.  
The two \td stars both have stellarity indices $>$0.9 and have FWHM nearly equivalent to the expected WFC3 PSF 
FWHM (0\farcs2).  Their surface brightness profiles are also indistinguishable from each other.
While the compact \zs\  galaxy  has a FWHM equal to that of the point sources, its surface brightness profile resembles
the other \zs\  candidate galaxy. This is   also supported by its lower stellarity  index of 0.6 and thus 
strengthens our earlier conclusion that this  object (zF25035) is a compact \zs\  LBG candidate.
Moreover, it shows that at \zs, there is a population of very compact (FWHM $\lesssim$1 kpc) star-forming galaxies
that are unresolved in typical image quality of ground-based imaging. 
Note however, that the FWHM or surface brightness profile can become unreliable at fainter magnitudes.
 
 To demonstrate this,  we have chosen a random star-like object   with stellarity $>0.9$ and FWHM similar  to those of the 
 compact galaxy, but with slightly fainter magnitude. As illustrated in  Figure~7 (grey dashed line), the surface brightness 
 profile of the random object is indistinguishable from galaxies despite its stellar-like FWHM.
% In our current case, SExtractor correctly identified the compact galaxy zF25035 as non-stellar, however if we were to 
% classify the random object (gray line) based on its stellarity or FWHM, it would be classified as a star despite  all the 
% similarities between the two.  
 This comparison demonstrates that the star versus compact galaxy classification based on either FWHM, stellarity 
 index, or even surface  brightness profile  would lead to mis-identification, especially at fainter magnitudes ($H \gtrsim 
 25$~mag), even with \hst-like image quality.
%
%This random object has a stellarity index $>0.9$ classifying it as a
%star.
%
In such cases, the near-IR \mb photometry will  allow us to cleanly distinguish between the spectral features of 
brown dwarf stars and \zs\  galaxy candidates.  

In summary, out of five  compact objects,  four   objects are  classified as brown dwarf stars, and the remaining  
object zF25035 is classified as a compact galaxy.   
Eliminating such objects due to their misidentification could affect  the bright end of the UV luminosity function  where objects are rare.  
This may  especially be of concern for  the case of  ground-based data when such compact galaxies are unresolved.

\subsection{Final Sample and New Selection Criteria using Medium-bands}

Using the initial color selection criteria   defined in Equation 1, out of  the 9 objects in the initial source list,  four objects are
identified as nearby  dwarf stars based on their FWHM and spectral template fitting test,  one object is  likely a foreground 
galaxy  based on  the S/N($>2$)  in the visible  $\chi^2$ image, and the remaining four  objects are   identified as \zs\   LBG
candidates.
To select the most robust \zs\  LBG candidates, we revise the   \zs\  candidate selection criteria using \mb filters as follows.
%in our sample, we further refine the color-color selection using the  \mb colors.

%Now
%that we have defined our \zs\  sample of three robust candidate
%galaxies, we analyze their properties and we measure their constraint
%on the UV luminosity function. 

%Based on the regions occupied by brown dwarf stars, a foreground galaxy, and \zs\  candidate galaxies in Figure~3, we derive 
%a  revised set of \zs\  candidate selection criteria as follows.

\begin{align}
 \mathrm {S/N }(J_{chq}) &> 7.0 \\
\mathrm {S/N}(\jtwo) & > 4.0 \nonumber\\ 
 \mathrm {S/N (\chi^2_{visible})} &< 2.0,\nonumber \\
  \jone -\jthree & < 0.4  \; \rm {mag},  \nonumber\\ 
  z' -\jone &  >0.8 \; \rm {mag}.\nonumber
%z'-  \jone &> 0.9 + 0.75(\jone - \jthree) \; \rm {mag}.  \nonumber
%z'- \jone &> -1.1 + 4(\jone - \jthree) \; \rm {mag}.\nonumber
  \end{align} 
This color-color selection is indicated in Figure~3 as the shaded region within the dotted polygon.   As can be seen, 
this color-color selection isolates \zs\  galaxy candidates from both low-redshift red galaxies and brown dwarf stars.    
Therefore, our new color selection criteria using the \FS \mb  filters  appears to provide a \zs\  LBG sample
that is free of contaminants,  
although future surveys (including spectroscopic confirmations) are needed to confirm this.  
The photometry of  all three \zs\  candidate galaxies is shown in Table 1, while Figure~4 and Figure~6 show postage stamps 
and best-fit SEDs obtained from  EAZY, respectively.

While this refined selection yields three robust \zs\  LBG candidates, one of the candidates (zF26343 in Figure~3) is excluded by this 
selection (which would have been selected if we had used the selection criteria from Equation 1), 
and will lead to  sample incompleteness thereby affecting the UV luminosity function.
We prevent this incompleteness by simulating the effective survey volume that is enclosed within the shaded region
in Figure~3.

%In the following sections
%we describe tests of the purity of this sample.  
%

\begin{table*}
  \centering
  \caption{Photometry of candidate galaxies at \zs.}
  \begin{threeparttable}
  \scriptsize { 
      \begin{tabular}{lccccccccccccc}
         \hline

 	 ID		& RA			& Dec  		& \zp	& \jone  		&  \jtwo		&  \jthree		&  \jhst		&  \hs		&  \hl 	 &	\hhst			&  \ks	&	$[3.5]$	&	$[4.5]$\\

	\hline
zF13965	& 	150.12579  	&    2.26662	&$>$27.4		&25.15		&25.24 		&25.32		&25.13 		&25.41		&25.25		&25.06 		&24.87  		& 24.04		&  24.80\\
		& 	  			&    		 	&	-		&$\pm$0.06	&$\pm$0.07	&$\pm$0.22	&$\pm$ 0.01	&$\pm$0.56	&$\pm$0.50	&$\pm$0.01	&$\pm$0.42	&$\pm$ 0.20	&$\pm$0.20	\\

zF25035	& 	150.09906   	&    2.34362	&$>$27.4		& 25.32		&  25.39		& 25.17		& 25.33 		&24.86		&25.47		&25.37 		&  25.24		& 25.58		&  23.64	\\
		& 	  			&    			&	-		&$\pm$0.06	&$\pm$0.06	&$\pm$0.15	&$\pm$0.02	&$\pm$0.30	&$\pm$0.56 	&$\pm$0.02 	&$\pm$0.45	& $\pm$ 0.50	&$\pm$0.20	\\

zF26836	& 	150.08658  	&    2.31785	&$>$27.4		& 26.39		& 26.09		& 26.28		&  26.68 		&27.09		&27.06		&26.47		&27.45		& $>$26.0	&$>$26.0	\\
		& 	 			&   		 	&	-		&$\pm$0.14	&$\pm$0.11	&$\pm$0.41	&$\pm$0.05	&$\pm$2.26	&$\pm$2.31	&$\pm$0.11	&$\pm$3.41	& -			& -	\\

zF26343\footnotemark[1]	&  150.07767  	&    2.35306	&$>$27.4	& 25.79		&  25.72		& 25.25		& 25.35 		&25.45		&24.91		&25.02		& 25.11		& 24.35		&24.42	\\
		& 	 			&   		 	&	-		&$\pm$0.11	&$\pm$0.10	&$\pm$0.21	&$\pm$0.02	&$\pm$0.65	&$\pm$0.43	&$\pm$0.05	&$\pm$0.51	&$\pm$0.20	& $\pm$0.20	\\

\hline

        \end{tabular}}
     \begin{tablenotes}
       \item[] All magnitudes are total magnitudes.
       \item[]   Limiting magnitudes ($2\sigma$ limits in 1\farcs1 diameter aperture): \gp=28.9;  \rp=28.9;  $i$=27.6. 
        \item[a] This candidate falls outside the refined color selection. \\
     \end{tablenotes}
  \end{threeparttable}
\end{table*}

\begin{figure}[t!]
\centering
\epsscale{1.2}
\plotone{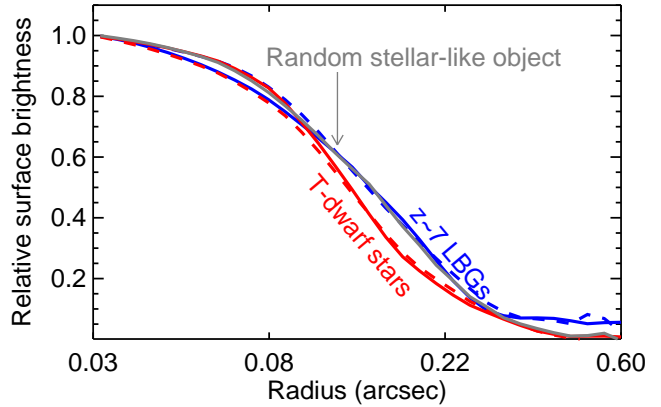}
\caption{Comparison of surface-brightness profiles (\jhst) of the compact \zs\  galaxy candidate (zF25035) and  point 
sources with stellarity index $>$0.9.  
Blue and red lines show surface brightness profiles of \zs\  LBG candidates and \td  candidates, respectively.
%Red lines show surface brightness profiles of  two objects selected as T-dwarfs,  while the blue solid line is the
%surface brightness profile of  the compact \zs\  galaxy candidate.
Dot-dashed line is a random object chosen to have similar magnitude and FWHM  as those of  the compact galaxy, except that 
this random object has stellarity index $>0.9$ identifying it as a star. 
 If we were to identify this random object based on its  FWHM or stellarity index it would be classified as a star, even though 
  its surface brightness profile resembles a non-stellar object. 
With typical ground-based image quality, and even with \hst\  image quality, 
 discrimination between galaxies and stars based on their stellarity index is unreliable 
   especially at fainter magnitudes.\\}
\end{figure}

% \begin{figure}[t!]
%\epsscale{1.22}
%\plotone{sel_sed_plot.eps}
%\vspace{-0.7cm}
%\caption{Spectral energy distribution of objects shown in Figure 7. Top four panels are  \zs\  LBGs, while the bottom panel represent one Tdwarf.
%Blue points represent \FS medium-band photometry.
%}
%\end{figure}

\section{Luminosity Function at $z \sim7$}
The UV luminosity  function provides a direct observational measurement  on the number density of star-forming galaxies at a given redshift.  
We derive the completeness-corrected luminosity function $\phi(m)$  as 
\begin{equation}
\phi(m)=\frac{1}{dm} \frac{N(m)}{V_{\rm eff}},
\end{equation}
where $N(m)$ is the number of detected objects with magnitude between $m$ and $m+dm$, and $\rm V_{eff}$ is the effective 
survey volume.

\subsection{The Effective Survey Volume}
In order to accurately estimate the observed luminosity function we must calculate the effective survey volume since this 
volume depends on both the apparent magnitude of the selected objects and their redshift distributions.  We estimate the 
effective volume $(V_{\rm eff})$ as described in   \citet{ste99}:
\begin{equation}
V_{\rm eff}(m)=\int \frac{dV(z)}{dz} p(m,z) \; dz,
\end{equation}
where $p(m,z)$ is a probability of detecting a galaxy of apparent magnitude $m$ at redshift $z$ specific to the image quality, 
depth, and passbands of our dataset, and $dV/dz$ is the comoving volume in a redshift interval $dz$.

In addition, due to  photometric errors,  objects can scatter in and out of the color-color selection region 
(shaded region; Figure 3).  In the following sections we describe how we estimate  the effective volume of our survey. 
In short, we first compute $p(m,z)$ by  inserting and then recovering  artificial galaxies in the real images as a function of 
apparent magnitude and redshift.  We then calculate $dV/dz$ for each redshift bin.

\subsubsection{Modeling the Colors of High-$z$ Galaxies}

To generate artificial galaxies that are representative of high-z galaxies, we created a grid of  spectral energy distributions (SEDs)  of star-forming 
galaxies using the  \citet{bru03}  models. 
In these models we used the Chabrier initial mass function, solar metallicity, and exponentially  rising star-formation 
history.   
For a given redshift, we simulated the effect of dust and IGM using the   \citet{cal97}  starburst  model  and the 
 \citet{mei06} IGM prescription, respectively.  
 We then convolved the resultant SEDs to calculate the bandpass magnitudes and colors through the   \zp\  and  \jone\ filters 
 at various redshifts  (see Figure 3). 
% We use these colors to calculate only the  \zp-\jone\ colors.  

To calculate the \jone-\jthree\ colors,  we followed a different procedure  instead of using galaxy tracks  
because the model  colors are  nearly constant (see Figure 3) over the redshift range we are currently probing.
Here  we  assumed that the flux at a given wavelength can be described as
$f_{\lambda} \propto \lambda^{\beta}$ where $\beta$ is the UV  continuum slope. For a given \jone\ magnitude (calculated from the
model SEDs as described in the previous paragraph), we calculate the \jthree\  magnitude assuming $\beta=-2.0$.  
This   value of $\beta$  is    consistent with $z\sim 7$ galaxies   \citep{fin11}.

\subsubsection{The Probability Function p(m, z)}
We computed the probability function (completeness function) by inserting artificial sources based on their redshifts and
 colors,  and recovering these artificial objects using the same procedure as for the real galaxies.  
 Now that we have magnitudes and colors of high-z galaxies as a function of redshift, we first have to create an artificial 
 object with some shape, and assign it a magnitude. 
 To create an artificial galaxy with its shape as close  as a real observed galaxy,  we choose a random  object in a given
 magnitude bin from the science image itself (in this case \jone). 

To insert this object in the  \zp, \jone, and \jthree\ images we choose 500  random positions avoiding any already existing 
sources ($> 3 \sigma$ flux) in the \jone\ image. Each object was  assigned  \ a \zp, \jone, and \jthree\  magnitude based 
on the slope $\beta$, and the color obtained from SED models,  respectively. 
We simulate artificial sources in a redshift bin of 0.2 and magnitude bin $dm=0.1$ mag.  
We then compute the recovered fraction by taking the ratio of recovered sources (using the selection criteria as 
described in Section 3.3) to the inserted number of objects.  
This completeness accounts for both the detection  incompleteness as well as the imperfect object selection due to
photometric errors leading to additional incompleteness.  
To increase  the number of simulated galaxies without increasing the crowding of sources in a single image, we repeat 
this procedure ten times for a given redshift and magnitude bin and then average the recovered fraction.

%To reduce the uncertainty due to lesser number of simulated objects we repeat this procedure ten times for a given redshift and magnitude bin
%and then average the recovered fraction. 
%
%\todo{We could show it for the brighter object, m=25 mag ? This might
%show our P(m,z) reaching higher peak (but the current figure is OK.}
As an illustration we show p(m,z) for \jone=24.8 mag  in Figure 8 (shaded area). To compare this completeness function
 to the \bb  filters, we also overplot the completeness function for the \zs\  selection using the $ F_{850}$, \yhst, and \jhst\  filters 
 from   \citet{oes12} and  \citet{bou12}.
 We clearly see a  broader completeness function for \bb  filters compared to our \mb  selection.  
 In contrast, at \jone=24.8 mag,   $p(m,z)$ is much lower than that derived from much deeper  \hst\  \bb surveys. 
Now that we have $p(m,z)$, we calculate the effective volume by integrating the product of $p(m,z)$ and $dV/dz$ 
(equation 5).

\begin{figure}[t!]
\centering
\epsscale{1.2}
\plotone{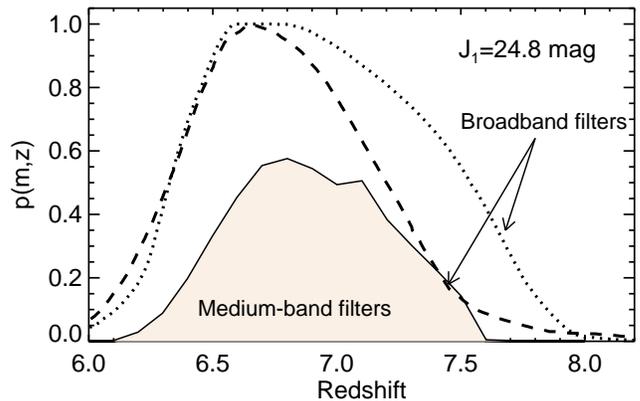}
\caption{ Probability of recovering artificial galaxies in our survey, shown for a source with \jone=24.8 mag (filled curve). 
% \zp-\jone\ colors are obtained based on the Bruzual \& Charlot models while \jone-\jthree\ colors are obtained using power law 
% slope defined as $ f \propto \lambda^\beta$ with  $\beta=-2$. 
 Dashed and dotted lines show \hst\ \bb completeness function from Oesch et al (2009) and Bouwens et al (2012),
 respectively.
  The  \bb completeness distribution achieves a higher peak $p(m,z)$   due to the \hst\ survey    depths,  but is  
  slightly broader   compared to the \FS completeness distribution.  The narrower distribution of medium-bands
allows for much tighter redshift probability distribution.\\}
\end{figure}

\begin{figure*}[t!]
\epsscale{0.9}
\plotone{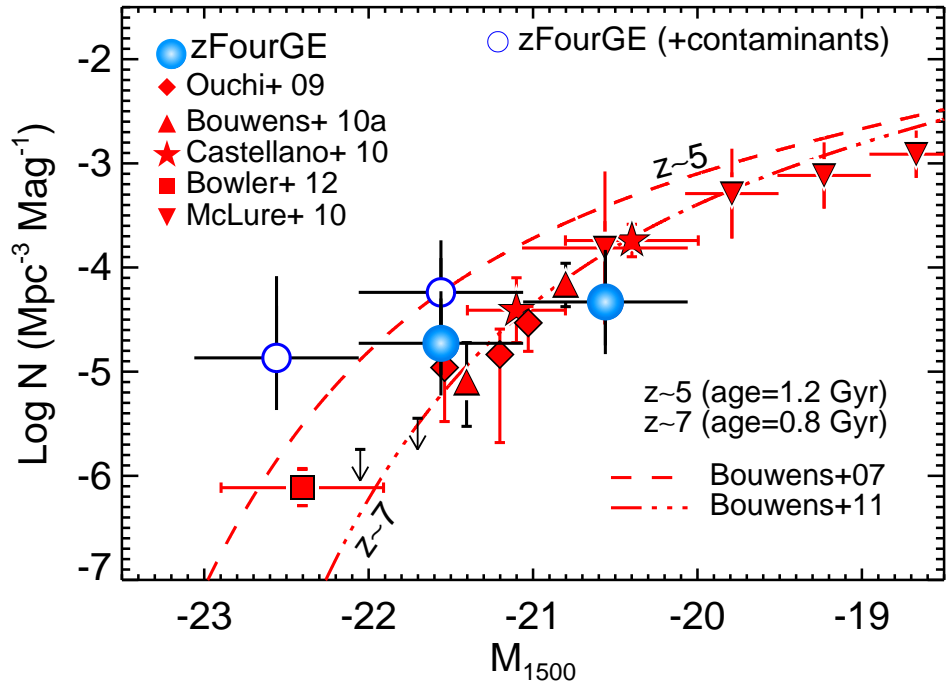}
\caption{ Binned UV luminosity function at $z\sim7$. 
Blue filled circles  represent the number densities of our \zs\  sample in   two rest-frame magnitude bins with $dM$=1 mag.
Error bars on the abscissa show the width of the magnitude bin.  
Error bars in the ordinate are a posteriori errors, calculated by marginalizing over all Poisson distributions given the observed 
number of two objects in the brighter bin and one object in the fainter bin.  
%These correspond to  Poisson errors only and do not include effects from, e.g., cosmic variance. 
The red colored filled diamonds, triangles, stars, squares, and downward pointing triangles  show luminosity functions from \citet{ouc09}, \citet{bou10a},  \citet{cas10}, 
\citet{bow12}, and \citet{mcl10}  respectively.  
The respective survey area for these studies are 1568 arcmin$^2$\citep{ouc09}, 58 arcmin$^2$\citep{bou10a}, 161 arcmin$^2$\citep{cas10},  3240 arcmin$^2$\citep{bow12},  and 
4.5 arcmin$^2$ \citep{mcl10}.
The dashed line and dot-dashed line show the best-fit Schechter functions at $z\sim5$  \citep{bou07} and \zs\  \citep{bou11} 
 respectively. 
Our  observations (blue filled circles) shows a moderate evolution at the bright end (\muv=-21.6 mag),  consistent with previous observations.
The open circles show our estimates if we  include foreground contaminants namely brown-dwarfs. Thus, confusing brown-dwarfs
for compact galaxies, which  is possible for ground-based broad-band observations,  will overestimate the UV LF.\\
 } 
\end{figure*}

%where
%$dV/dz$ is the comoving volume per square arcminute in redshift slice dz at redshift z and Az	is the area of the field in arcminutes.

\subsection{Evolution of UV Luminosity Function at $z \sim7$}
The apparent magnitude can be converted to the  rest-frame UV magnitude $M_{UV}$ (at $\lambda=1500\AA$) using
\begin{equation}
M_{1500}=\jthree -DM(z) -k,
\end{equation}
where  DM is the  distance modulus  and  $k$ is the $k$-correction between the rest-frame wavelength 1500\AA\  and  \jthree\  
filter.  Assuming a negligible $k$ correction, the above equation can be rewritten  as 
\begin{eqnarray}
M_{1500}=\jthree - 5 \; {\rm log}(D_L/1 \;pc) +2.5 \; {\rm log}(1+z) +5,
\end{eqnarray}
where $D_{L}$ is the luminosity distance at redshift $z$.

We divided our \zs\  sample in two magnitude bins,  with each bin 1 magnitude  wide. The brighter  luminosity bin contains two objects, 
while the fainter bin contains the remaining one object.  Figure 9 shows the UV luminosity  function of $z\sim7$ galaxies derived from our 
observations for these three galaxies (blue filled circles).  

 Currently there are only a few   other studies exploring the UV luminosity  function at
magnitudes brighter than \muv = -21.5 mag 
 \citep{ouc09,cap11, bow12, hat12}.
  Our result is consistent with  previous conclusions
about the evolution of the UV luminosity function:  the UV luminosity  function evolves  from $z\sim
5$ to $z\sim7$ at the bright end ($M_\mathrm{UV}\sim-21.5$).   
 Thus,  within $\sim$ 400 Myr the
number density of $M_\mathrm{UV}=-21$ to $-22$  mag galaxies has
increased by a factor of $\sim4$ from \zs\ to $z\sim5$, which implies a
rapid build-up of the star-formation rates in galaxies during this
short period.

\subsection{Comparison to other \zs\  Samples}

While there are a few other studies focused on \zs\  LBGs using \hst\ observations, currently there are only four  other
ground-based surveys that have identified \zs\  dropout galaxies using \bb  imaging (Figure 10: 
 \citealp{ouc09,cas10,bow12, hat12}). 
  Using Subaru/Suprime-Cam in the Subaru Deep  and GOODS-N  fields,   \citet{ouc09}  identified 22 \zs\  LBGs down to
  Y=26 mag.  
  Our current survey reaches a similar depth
with \jone =26.1 mag (this filter is similar to the Y filter),  but   
  covers a smaller survey area.   \citet{cas10}  found eight $z$-dropout galaxies in the GOODS-S field using HAWK-I  
  and  FORS2 \bb  (Z, Y, J)  observations in a survey area  comparable  to ours, however with shallower
  survey depth.
%\todo{how does
 % their depth compare to ours?}  
  Recently,   \citet{bow12}   found about ten \zs\  LBG candidates in a large survey area (UltraVista) in the COSMOS field. While their survey depth is shallower (brighter than $J$=25 mag), their survey area is much larger allowing them to probe 
  the brightest and rarest galaxies at this redshift.
Finally,  \citet{hat12} found  two possible LBGs brighter than $J$=24.5 mag from a slightly wider ($\sim169$ arcmin$^2$) 
but much  shallower ($J\sim25 ;3\sigma$) survey  in the GOODS-N field. 

On the other hand, using \hst\ observations, several studies  have found \zs\  dropout galaxies, all fainter (\muv $\gtrsim 
-21.5$ mag) than the brightest galaxies  from   ground-based observations.  
All these studies use primarily three fields: HUDF  \citep{bou11,gra11},  GOODS-S  \citep{wil10,gra11}, or GOODS-N 
\citep{hat12}.
 While deep \hst\ observations are appropriate in probing the faint end of the UV luminosity function, ground-based surveys have the 
 advantage of probing much larger volume and thus finding brighter and rarer galaxies, allowing us to constrain the bright  
 end of the luminosity  function as can be seen in Figure 9.  
 All the observations brighter than \muv = -21.5 mag are currently obtained from ground-based observations.

From Figure 9 we can see that at brighter magnitudes (\muv\  brighter than  -22 mag), the best-fit  Schechter functions
 obtained for  the HST observations   \citep{bou07, bou11} prefer a significant evolution from $z\sim5$ to \zs.
On the other hand, the ground-based observations \citep{ouc09,bow12}  suggest a less dramatic evolution at the bright 
end, however due to large error bars, strong evolution can not be ruled out.
It is possible that  in the absence of a clean method to identify contaminants, namely T-dwarfs (square symbols in Fig. 3),
the observed \zs\  UV luminosity function that includes contaminants (white filled circles in Fig. 9)  will be  overestimated  and consistent with 
$z\sim 5$ luminosity function \citep{bou07}. This is especially likely to be the case for the ground-based broad-band
imaging where atmospheric seeing makes it difficult to distinguish between stars and unresolved galaxies based 
on their FWHM; \zf medium-band filters provide an effective   method to identify T-dwarfs.
Based on current observations, while the bright end of the UV luminosity  function at \zs\  is reaching a better observational consensus, 
larger-area surveys with improved 
sensitivity are required to improve the UV luminosity  function measurement and   test for cosmic variance.

%\subsection{Expected Number of \zs\ LBGs in Our Survey}
%While we have identified five \zs\ LBG candidates from our survey, we expect about 20 \zs\ LBGs assuming 
%observed UV LF from Castellano et al (2010). [Check: This number assumes Veff=1.0D5]

\section{Physical properties of $z\sim7$ Galaxies}

%Samples of candidate LBGs at \zs\  have been steadily increasing,  and our understanding of these objects has improved 
%significantly over the last few years.  
%Nevertheless,  their faintness  makes it difficult to fully understand the physical  nature of these objects.  
Because  near-IR \mb filters  provide a higher spectral resolution  compared to the \bb  data,   the constraints on physical
parameters of  these galaxies is improved (photometric redshifts, stellar masses, star-formation rates (SFRs), ages, extinction, etc.).  
Here, we constrain the physical properties of the two brightest galaxies, which have high S/N in the \zf  bands.  
%
%An additional
%uncertainty on the derived physical properties can be attributed to
%the broadband  photometry.  This is because, due to the wide wavelength
%range of the broadband filter, there is a large uncertainty  in the
%photometric redshift which will have  a measurable  effect on the
%derived physical properties.

%As we demonstrate in the following sections, some uncertainty can be
%improved using photometry obtained from MB filters.  In particular, 
It is  standard practice to compare  galaxy photometry with stellar population models that vary over some range of 
parameter space.
Uncertainties on the stellar population parameters (e.g., mass, age, extinction, star-formation history, metallicity) are 
either derived through Monte Carlo methods or by marginalizing over some probability distribution function.  
Both of these methods assume a prior that the  models represent the data.  
%As we will show below, the addition of the higher-spectral resolution MBs shift the stellar population constraints into 
%regions disfavored when only \bb filters are available.   
%Therefore, including the MBs allows us to measure more accurately the constraints on the stellar populations. 
In the sections that follow we compare the physical properties derived using  \bb  photometry alone, and then including 
the  \mb  photometry. 
%
% In this and the following sections we  focus primarily on the
 %two  brightest galaxies due to their high S/N detections.
%
 For illustrative purpose, we plot  the relevant figures either for the brightest (zF13965) or the 2nd brightest (zF25035) LBG
 (whichever yields  extreme values) but  tabulate  the derived   properties for the  two brightest  galaxies in Table 2-3.

 \subsection{Stellar Population Synthesis Modelling}

 To derive  the physical properties of our $z\sim7$ galaxies, we obtained  model stellar population spectra using the  
 \citet{bru03}   stellar population synthesis code.  
 We generated model spectra using  a grid of dust,  star formation histories (SFHs), and stellar population ages (t) 
 ranging from 10 Myr to 750 Myr  in logarithmic intervals.
These limits on stellar age  ensure that it is greater than the expected dynamical timescale of the galaxies but less than the age of the 
universe at a given redshift.
We restrict  the metallicities to Z=0.2\zsol\  and assumed a Chabrier initial mass function (IMF) -- for a Salpeter IMF the 
masses would be  roughly 0.2 dex larger, but this would have no effect on the rest-frame UV-optical colors of the
galaxies as the slopes of the IMFs at the high-mass end are identical between the two IMFs.
In addition, we included nebular emission lines in our synthetic model spectra. 
These lines have been shown to affect the derived physical properties of galaxies  in the local universe 
\citep[e.g.,][]{pap02, pus04}  and at high-redshifts \citep[e.g.,][B. Salmon et al 2013 in 
preparation]{sch09, sch10, fin12}.
We constrained $\tau$\footnote{We model the star-formation rate  as a function of  $e^{-t/\tau}$.}
 to be negative (rising SFH), constant,  or $\tau>$300 Myr as the SFHs of these galaxies are expected (on 
average) to be increasing with time \citep{pap11}.
%
% and we constrain the stellar age to be
%greater than the likely dynamical timescale ($>$50 Myr) at \zs.
Next,  we applied the dust attenuation to our model spectra using the Calzetti  dust law  \citep{cal00}, which is 
appropriate  to starburst galaxies. 
The spectra were  then redshifted and attenuated using the IGM attenuation from  \citet{mei06}. 
Finally, we computed the bandpass averaged fluxes in each of the filters.  
The best-fit model was  obtained by minimizing the $\chi^2$ between the bandpass averaged fluxes and the model 
spectra. 
To understand the effect of different dust attenuation laws on the SED-derived  parameters, we also used
the SMC-like dust law \citep{pei92} described in Section  5.3.

\begin{figure}[t!]
\epsscale{1.16}
\plotone{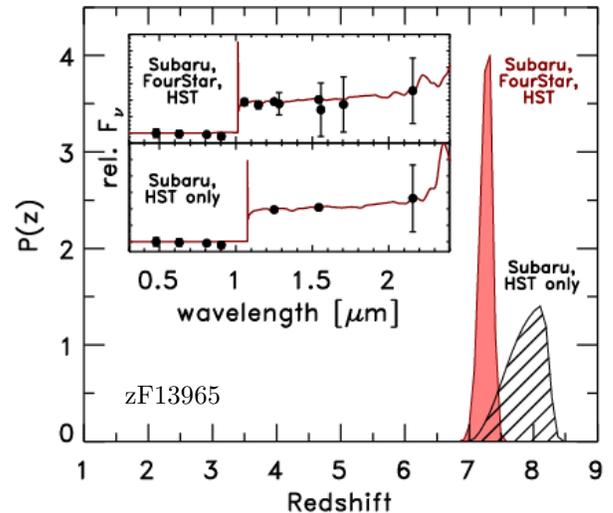}
%{\bf  {\tiny\put(-182,177){IRAC}}}
%{\bf  {\tiny\put(-182,133){IRAC}}}
%{\bf  {\tiny\put(-45,177){IRAC}}}
%{\bf  {\tiny\put(-45,177){IRAC}}}
{\normalsize \put(-190,45){zF13965}}
%\vspace{-0.3cm}
%\epsscale{1.06}
%\plotone{pzFSvsHST_27761.eps}
\caption{Comparison of photometric redshift probability distribution of the brightest object (zF13965) using only broad-band photometry  (\hst +Subaru+IRAC: hashed region),  and adding \mb filters (\FS + \hst +Subaru +IRAC bands : filled red region).  
The photometric redshifts using the \FS \mb filters   prefers a lower as well as narrower redshift range. \\}
\end{figure}

  \subsubsection{Photometric Redshift  Distribution}
Figure~10 shows the photometric redshift probability distribution for the  brightest object.
For the  brightest object (zF13965),
the photometric redshifts derived for the \hst+Subaru bands tend to be higher with  $z_{\rm phot}$=8.11$^{+0.45}_{-1.06}$, 
while inclusion of \mb  photometry lowers the photometric redshift to $z_{\rm phot}$=7.24$^{+0.38}_{-0.25}$ (where the 
error bars are all 99\% confidence; Figure 10).  This is also true for the 2nd brightest object (zF25035),
with $z_{\rm phot}$=7.66$^{+0.77}_{-0.57}$ and $z_{\rm phot}$=7.16$^{+0.35}_{-0.19}$ 
when derived using \bb  photometry alone, and then adding \mb filters, respectively.   
The addition of \mb  photometry lowers the median redshift  because of the \jone\ filter, which better constrains the 
Lyman-break.  
Surveys using a Y-band \nir  filter may  expect similar photometric redshifts.  
Including the \mb filters, the photometric redshift probability distribution functions favor lower redshift solutions, with a 
narrower range of favored redshfits.  
%This is illustrated by the 99\% confidence regions on the photometric redshifts above.     
We fit the stellar-population models to the measured photometry over a range of redshift spanned by these
probability distribution functions, and we marginalize over these properties when determining constraints on stellar 
population parameters.

\begin{figure}[t!]
\centering
\includegraphics[scale=0.57]{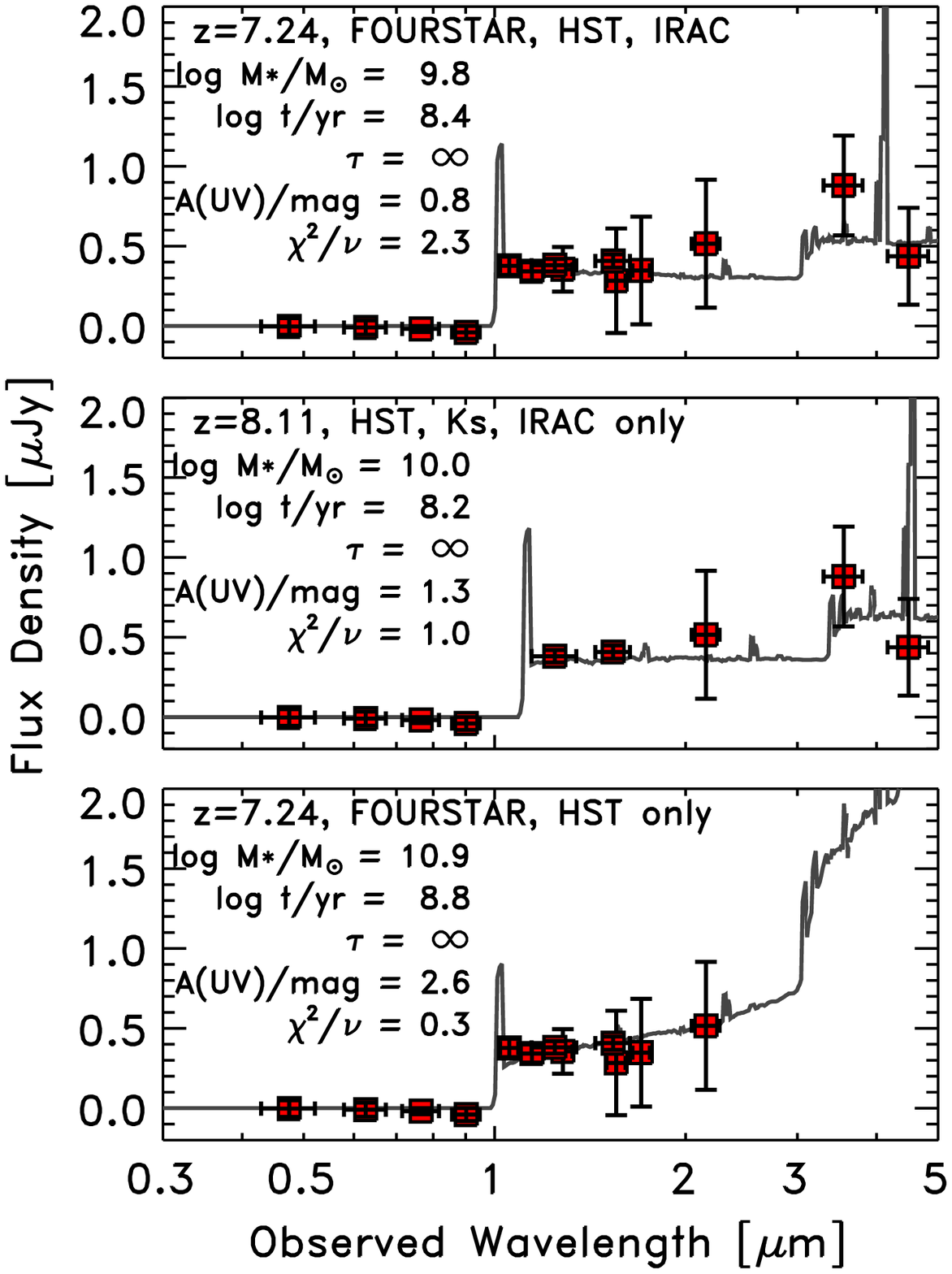}
{\normalsize \put(-80,295){zF13965}}
\label{fig3.1}

%\vspace{-7.8cm}
%\hspace{4 cm
%\includegraphics[scale=0.6]{beta_slope_plot.eps} 
%}

%\vspace{5.8cm}

\caption{ Comparison  of SED of our brightest candidate (zF13965) using all bands (top panel: \hst+Subaru + \FS \mb+IRAC)
 and excluding \mb  photometry (middle panel: \hst+Subaru +IRAC bands).
 The SED model assumes 0.2\zsol\ metallicity, a Chabrier IMF, the Bruzual \& Charlot 2003 model, and include nebular
emission lines. 
The lower panel shows best-fit SED without IRAC bands.
Adding \FS   \mb filters  improves the constraints on the photometric redshift  and other physical properties,  while exclusion of 
IRAC bands yields poorest constraints.\\}
%In this case the break appears between the z and J1 bands.  The z - F125W allows too much room in z-space.}
\end{figure}

 \subsubsection{Best-fit SED}
Figure 11 shows the best-fit model spectrum with and without the \FS \mb   filters for our brightest \zs\ LBG.  
The top panel shows the best-fit model spectrum using  \FS \mb filters, \hst\, and visible bands while the middle  panel  shows 
the best-fit SED without the \FS \mb  data.  
The bottom panel shows the effects when the IRAC bands are excluded.  In all three cases, we use the photometric
redshifts obtained from EAZY, using respective bands only. 
The behavior of the best-fit SEDs for the second brightest \zs\ galaxy  are very  similar.

 \subsubsection{UV Spectral Slope}

The UV continuum light is sensitive to dust absorption and its  slope  provides a good tracer of the dust 
attenuation for star-forming galaxies as the two are well correlated in UV-luminous, star-forming galaxies from $z\sim 0$ to 
$\sim 3$  \citep{meu95,meu97,meu99, dad04,bur05,lai05,red06,dal07,tre07,sal07}.

 Recently,  \citet{fin11}  measured UV spectral slopes of \zs\ galaxies  by directly fitting the best-fit SED with a power-law.  
 %They found that at $z\sim4$, the values of $\beta$ obtained by measuring $\beta$ directly from the best-fit stellar population 
 %spectrum  and a single color differ significantly.  
% On the other hand, at higher redshifts (\zs), the two values are comparable.  
 We followed a similar procedure to measure $\beta$  by fitting a power-law to the best-fit SED in the  the  rest-frame 
 wavelength range  $1250 \le \lambda \le  2600$ \AA\  \citep{cal94}.

For the brightest LBG in our sample (zF13965),   with this method we find $\beta$=-1.88$\pm0.1$ using  the \hst\ bands while adding  
the   \mb photometry gives a steeper value of $\beta$=-2.08$\pm0.1$  (see Table 3). 
%
% which is somewhat opposite to what we got using the
%single color method (equation 8).
%
For the 2nd brightest LBG  (zF25035),   we find a similar trend in $\beta$ with a shallower  slope,  $\beta=-1.29$$\pm0.05$  using  \hst\  \bb 
photometry and a steeper    $\beta=-1.44$$\pm0.12$  when adding \mb  photometry.  
For  both galaxies, our analyses favor steeper  (i.e., \textit{bluer})  UV spectral slopes when the \mb  photometry are included.
 
The UV slope measured for both objects  are  between  $\beta$=-1.2 to $\beta$=-2.1,  within the range seen in other studies at 
\zs.
Finkelstein et al (2011) found a median value of $\beta$=-2.1 for galaxies with $\rm  M_{UV}=-20.6$.  
Similarly, Dunlop et al (2012) found an average value  of $\beta$=-2 for bright \zs\  galaxies with $\rm M_{UV}=-20.5$. 

 \begin{figure}[t!]
\centering
\includegraphics[scale=0.54]{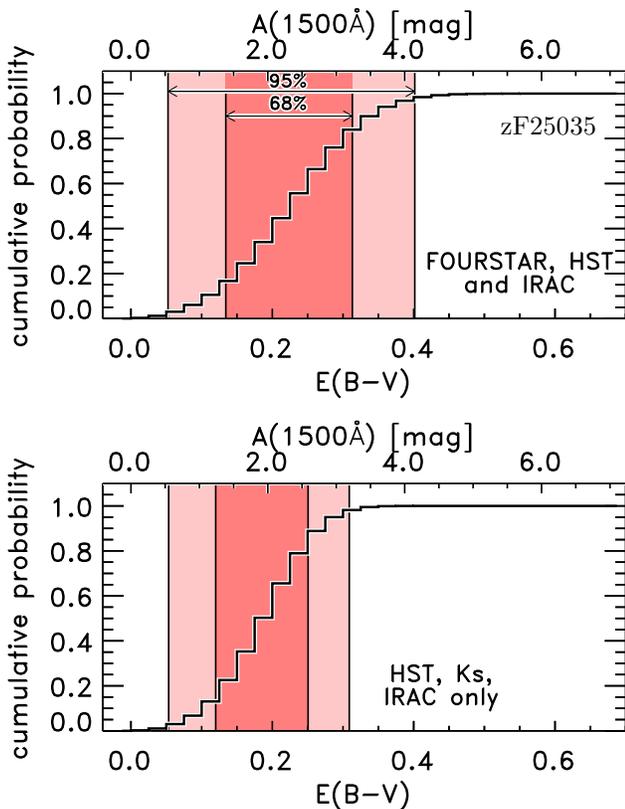}
{\normalsize\put(-60,260){zF25035}}
%\put(-80,80){zF25035}
\label{fig3.1}
\vspace{-0.6cm}
\caption{Cumulative probability distribution of  E(B-V), for the  $2^{nd}$ brightest candidate (zF25035), using \bb photometry alone 
(top panel), while the bottom panel
includes \mb photometry. 
While the median E(B-V)=0.2  are nearly same for both,  the E(B-V) derived using  \mb filters  allows a wider  distribution compared to the
the distribution obtained using \bb photometry alone.\\}
%The median E(B-V)  derived using \bb photometry alone is somewhat lower with E(B-V)=0.01 compared
%with E(B-V)=0.09 derived after adding \mb photometry.

\end{figure}

   \begin{table*}[t!]
  \centering
  \caption{Physical properties derived from spectral energy distribution fitting   for the  two brightest 
  LBG candidates (zF13965 \& zF25035).}
  \begin{threeparttable}
  \small{ 
      \begin{tabular}{l l c c c c c c}
\hline
         
 ID		&	Filters 	& \zphot	& $\beta$		&	E(B-V)\footnotemark[1] 	&  Log Stellar  	&  Stellar age\footnotemark[2] & Best-fit \\
		& 			& 	 	 &			&    						 &mass \msun 	&	Myr					& $\tau$ (Gyr)\\
\hline

zF13965	&	\hst+Subaru+IRAC+\FS			& 7.24$^{+0.38}_{-0.25}$  	&   -2.08$^{+0.10}_{-0.10}$		& 0.09$^{+0.06}_{-0.05}$			&  9.8$^{+0.2}_{-0.2}$	& 251		& 100 		\\
		&	\hst+Subaru+IRAC				& 8.11$^{+0.45}_{-1.06}$  	&   -1.88$^{+0.10}_{-0.10}$		& 0.01$^{+0.20}_{-0.01}$			&  10.0$^{+0.2}_{-0.3}$	& 159 		& 100 \\
		&	\hst+Subaru++no IRAC+\FS		& 7.24$^{+0.38}_{-0.25}$ 	&   -1.17$^{+0.05}_{-0.05}$		& -							&  11.3$^{+0.5}_{-0.7}$	& 630		& 100\\
\vspace{0.1cm} \\	
\cdashline{1-7}

zF25035	&	\hst+Subaru+\FS				& 7.16$^{+0.35}_{-0.19}$  	&   -1.44$^{+0.12}_{-0.12}$		& 0.22$^{+0.09}_{-0.09}$			& 10.4$^{+0.3}_{-0.4}$	& 321	  	& 100	\\
		&	\hst+Subaru					& 7.66$^{+0.77}_{-0.57}$  	&   -1.29$^{+0.05}_{-0.05}$		& 0.24$^{+0.07}_{-0.20}$			&  10.2$^{+0.3}_{-0.3}$	& 640 		& 100 \\
		&	\hst+Subaru++no IRAC+\FS		& 7.16$^{+0.35}_{-0.19}$  	&   -1.52$^{+0.10}_{-0.10}$		& -							&  10.4$^{+0.5}_{-0.4}$	& 640	  	& 100	\\

\hline      
        \end{tabular}}
     \begin{tablenotes}
      \scriptsize { 
       \item[a] Median value.  Errors indicate upper and lower 68\% values.
%        \item[ ] Errors indicate upper and lower 68\% values.
         \item[b ] We report the best-fit ages from the spectral-energy distribution modeling only, as the age constraints depend strongly on the assumed 
         star-formation history. 
        }
          \vspace{0.5cm}
     \end{tablenotes}
  \end{threeparttable}
\end{table*}

\subsubsection{Constraints on Dust Attenuation from SED Fitting}
\citet{fin10}  found the best-fit A(V)=0 $-$ 1.5 for a sample of LBGs at \zs\  obtained from the \hst\ WFC3 in the HUDF field, while
\citet{lab10}  found A(V)=0 for a stacked sample of \zs\  LBGs, also obtained from the \hst.   
Figure 12 shows  the cumulative probability distribution of the dust content E(B-V) for our 2nd brightest object (zF25035)   from the
 SED fitting.  
 The shaded regions show 68\% and 95\% confidence regions. 
 The  top panel  shows E(B-V) with \bb+\FS   photometry, while the  bottom panel shows the probability when including \bb 
 photometry alone (Table 3). 
%  In both the plot and the inset, the lower end of the 68\% and 95\% confidence regions both lie in the first bin at E(B-V)=0.0.  
%  This is an artifact of the binned cumulative distributions, where more than 16\% of the cumulative distribution lies in the bin at 
 % E(B-V)=0.0. 
% The \mb\ photometry allows much wider distribution of E(B-V) compared with the \bb\ photometry alone (bottom panel). 

 While the best-fit model,  with and without the medium-bands,  yield similar range of E(B-V), the    probability distribution with the 
medium-bands spans a wider range  with a median  E(B-V) $=$ 0.22, slightly lower compared with the median E(B-V) that 
excludes the medium-band filters.
%On the other hand, in the inset  plot the median E(B-V) $=$ 0.01 but  spans  a much  wider range of E(B-V). 
% In part this is because we have adopted the physical prior that E(B-V) be non-negative.  
%While the constraints from the \hst\ data alone favor lower E(B-V), they can not have \textit{negative} E(B-V), and therefore the
%distribution is truncated.   
%The fits with the \FS MBs favor slightly higher E(B-V).  
%In both cases, fits with E(B-V)=0.0 are still allowed, but the distribution extends to higher values.    
For the brightest LBG (zF13965), median E(B-V)=0.01  and E(B-V)=0.09 with and without \FS bands respectively.
For the reasons described above, the \FS bands provide a more robust constraint on the physical 
properties of galaxies owing to their higher  spectral resolution.

\subsubsection{Stellar Mass and  Age}

Bowler et al.\ (2012) found that  LBGs at \zs\  with magnitudes in the range of our two most-luminous sources ($J\sim25$ mag)
have stellar mass $\sim 5 \times 10^{9}$\msun.  
For fainter magnitudes, depending on the range of extinction, the estimated stellar masses range from  $3\times  10^{7}$ - 
$1\times 10^{10}$\msun  \citep{fin10, lab10}.

\begin{figure}[t!]

\epsscale{1.14}
\plotone{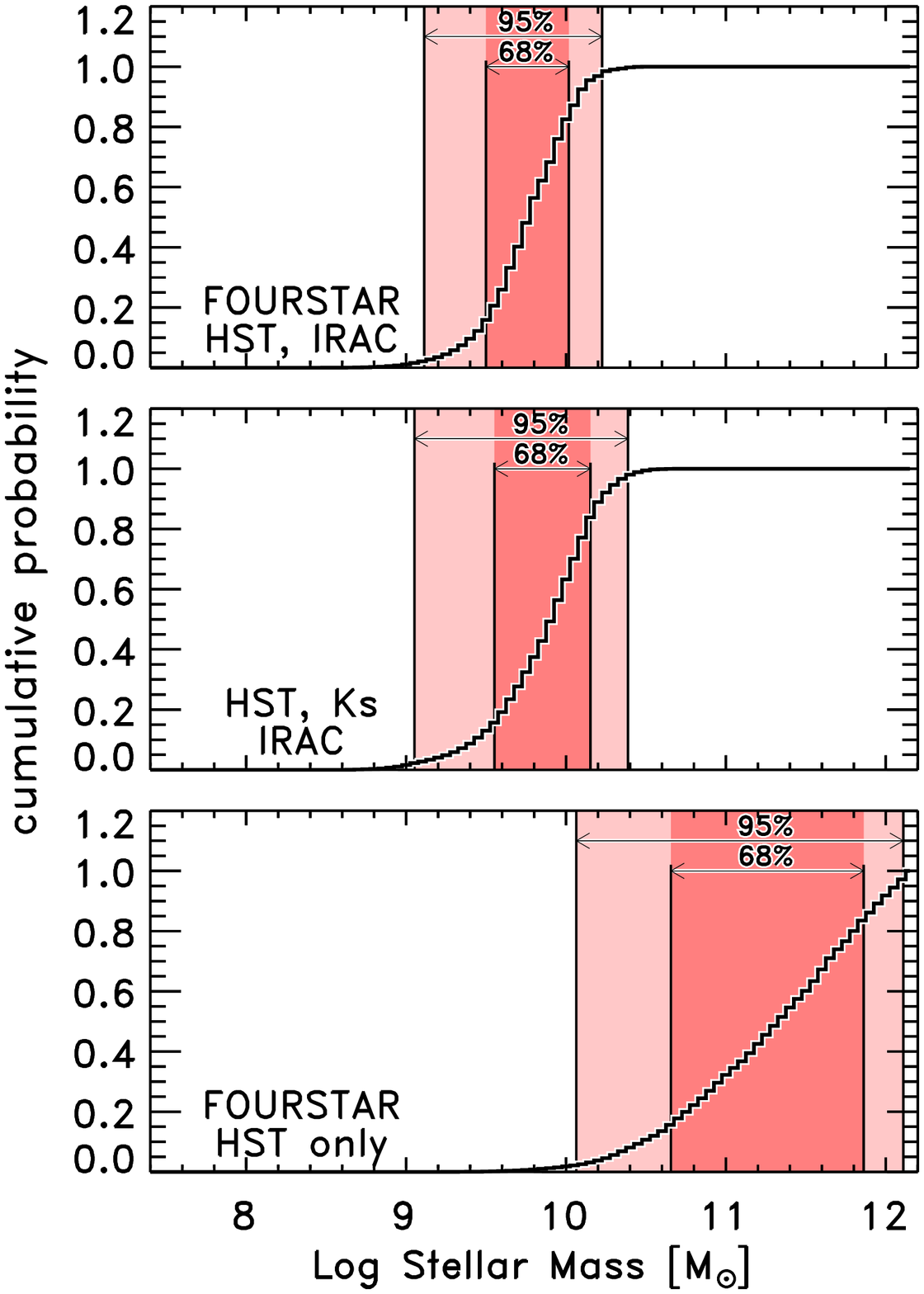}
{\normalsize \put(-180,300){zF13965}}
%{\normalsize \put(-180,205){zF13965}}
%{\normalsize \put(-180,110){zF13965}}
%\vspace{-0.8cm}
\caption{ Cumulative distribution of stellar mass for the most luminous \zs\  LBG (zF13965) with all bands included (top panel),
excluding \FS \mb filters  (middle panel), and excluding IRAC bands  (lower panel).
The addition of \mb photometry shifts the  stellar mass distribution  to  lower values  (from $ 1\times 10^{10}$~\msol\ to 
$6.3\times 10^{9}$~\msol) because it better measures the spectrum of the galaxy.
Removing the IRAC data allows for a large range of stellar-population parameters, including older age models, which greatly 
extends the range of stellar masses to higher values.  \\
%The second brightest \zs\  LBG (zF25035) exhibits very similar behaviour in its cumulative distributions.  
%This illustrates that adding additional bands shifts the allowed regions of the stellar-population parameter space,
%improving our physical understanding of the galaxies.\\}
%
}
\end{figure}

For our brightest   object (zF13965),  the favored range of stellar mass shifts to lower  values  after including the \mb  filters, with the
 most likely median mass decreasing  by a factor of 1.5.
 % for both of the luminous \zs\  galaxies in our sample. 
 As can be seen in Figure  13, the probability distribution is narrower with median mass $ 6.3\times 10^{9}$~\msun\
 when the \mb filters are included;  the absence of the \FS \mb photometry shifts the masses to higher  values (median mass  $= 1\times 10^{10}$~\msun).
 % \msun), and this increases to $ 1.6\times 10^{10}$~\msun\ when the MBs are included. 
On the other hand, for the second brightest object (zF25035), the median mass is higher ($ 2.5\times 10^{10}$ ~\msun) 
when  \mb filters are  included.
 % The difference can be attributed to the low dust content in the best-fit SED model when only the \bb  photometry is 
 %included -- the best-fit SED using only the \hst\ data favors bluer rest-frame UV colors which drives the fits to models with 
 %lower mass-to-light ratios.    
%Adding the \mb filters improves our ability to constrain the physical parameters in these galaxies, including quantities such as  mass.

Figure~13 (bottom panel) shows the constraints if we remove the IRAC bands. 
In this case, the allowed range of stellar population  parameters that fits the rest-frame UV extends to models with very old
ages and high mass-to-light ratios, and the fits are no longer able to exclude these possibilities.    
Thus, for the brightest object, the IRAC photometry is essential  to better constrain the best-fit stellar masses.
 % \todo {Include actual numbers without IRAC. We should include a sentence here about the effect of not including the Lya line or excluding
 %the J1 filter}.
 For the second  object however, the presence of IRAC bands does not change the derived stellar mass (Table 3).
  A likely reason for this is that in our models the redshifted [OIII]  line flux falls in the 4.5$\mu$m  band over the photometric redshift range.  
 This boosts the flux in the model, and requires lower stellar masses.
 
 The stellar ages for the two brightest objects  range from 250 to 640 Myr, albeit with large uncertainties which depend
 on the assumed star-formation history.
 We can however estimate the timescale over which the stellar mass doubles. Because the best-fit $\tau$ (Table 3) for 
 both  objects is large (100 Gyr), the timescale is simply stellar mass/SFR. This yields a timescale of about 150 Myr to double 
 the stellar mass.

In summary, using  \bb   photometry alone,  broadens the  photometric  redshift range compared to the result when the \mb  data
 are included.
This is expected as the \mb filters  more accurately isolate the location of the Lyman-break. 
In addition, currently there are no \bb filters corresponding to the central wavelengths of some of the \FS \mb filters (see Figure~1).

%\textbf{ We note that increasing the number of 
%filters will yield  a more accurate best-fit SED, but the higher spectral resolution of medium-band filters will have a greater impact.}    
%Moreover, using only the \bb  photometry places constraints on the allowed range of stellar population parameters that 
%appear too severe compared to the allowed range on the parameters when the MBs are included.   
While our comparison is based on only two \zs\  dropout galaxies,  they are at the upper luminosity (and stellar mass) range of
 galaxies at this redshift.  
For our small sample, adding the \mb photometry to the SED fitting yield more accurate constraints on the physical properties of 
the \zs\  galaxies compared to using \bb photometry  alone.  To generalize these conclusions to the \zs\   galaxy population will 
however require larger samples with deep \mb  imaging.

 \subsubsection{ Effect of \lya\ Line and \jone\ Filter on SED Fitting}
 
 We further investigated the presence of a \lya\ line and the effect of  the \jone\ filter on the derived physical properties of our bright LBGs.
 To do this, we  obtained the best-fit SED (section 5.1)
 by first including no \lya\  emission and then by excluding the \jone\  filter.
    Excluding the \jone\ filter or absence of a \lya\ flux results in essentially the same values as that for \bb only derived parameters.
This reinforces the idea that we need \lya\ equivalent width  measurements to understand  (1) the dust extinction and (2) the stellar masses, 
especially for galaxies with \lya\ emission line.

  \subsubsection{\lya Equivalent Width}
In principle,  the \lya\  equivalent width $W$ provides a proxy to the star-formation rate in the  galaxy, and  defined as 
 $W \equiv F_{\rm line} / F_\lambda^c$, where $ F_{\rm line}$  is the total line flux (\ergscm) and $F_\lambda^c$
 is the continuum flux density (\ergscma).
 If our \zs\  LBG candidates have \lya\ line, it will contribute a flux excess in the \jone\ filter, thereby decreasing (i.e., making
 brighter)  the \jone\  magnitude \citep{pap01};
 \begin{equation}
\Delta m \simeq -2.5 \log\left[ 1+\frac{W_0 (1+z)}{\Delta\lambda}
\right], 
\end{equation}
where  $W_0$ is the rest-frame \lya\ equivalent width and $\Delta \lambda$ is the \jone\ filter width.

To measure  $\Delta m$, we followed two different methods to take advantage of the presence of \jone\ filter.
In the $\rm 1^{st}$ method, we first obtain  the best-fit SED using all photometry and including the emission line 
(see  Section 5.1). We then use this
best-fit SED, but remove the emission line, and measure the bandpass \jone\ magnitude.
This magnitude compared with the observed \jone\ magnitude yields $\Delta m$.
In the $\rm 2^{nd}$ method, instead of removing the emission line, we obtain the best-fit excluding the 
\jone\ filter and compare the observed \jone\ magnitude with the synthesized \jone\ magnitude to get $\Delta m$.
We calculate the rest-frame equivalent widths using Equation 8 for our two brightest \zs\   LBG candidates (Table 4).
Using these $W_0$, we calculate the \lya\ line flux: 
\begin{equation}
F_{\rm line}= F_\lambda^c \;  W_0 \;  (1+z).
\end{equation}
 \begin{table}[t!]
  \centering
  \caption{Estimated rest-frame \lya\ equivalent widths and \lya\ line fluxes for the two brightest \zs\  LBG candidates.
  }
  \begin{threeparttable}
  \small{ 
      \begin{tabular}{lcccc}
\hline
 ID		&	&	EW$_0$ 	&	&	\lya\ flux \\
		&	&	(\AA)	 	&	&	$(10^{-17}$ \ergscm)\\
 % \hline
%zF13965	&	&	16.5		&	&	1.15		\\
%zF25035	&	&	26.2		&	&	1.07		\\

\hline      
\textit{Method 1} &	&							&	&							\\
zF13965	&	&	16.5$^{+15.3}_{-15.3}$		&	&	1.15$^{+1.07}_{-1.07}$		\\
zF25035	&	&	26.2$^{+16.1}_{-16.1}$		&	&	1.07$^{+0.66}_{-0.66}$		\\
\hline      

\textit{Method 2} &	&							&	&							\\
zF13965	&	&	-1.0$^{+12.6}_{-12.6}$		&	&	0.0$^{+1.0}_{-1.0}$		\\
zF25035	&	&	13.5$^{+14.3}_{-14.3}$		&	&	0.6$^{+0.66}_{-0.66}$		\\
\hline

\vspace{0.2cm}

        \end{tabular}}
  \end{threeparttable}
\end{table}
Table   4 shows the estimated \lya\ EWs and line fluxes, obtained from both methods, for the the two brightest \zs\  LBG candidates.
The  \lya\ EWs are smaller when the \jone\ filter is excluded ( $\rm 2^{nd}$ method). On the other hand, 
the  \lya\ fluxes obtained from the $\rm 1^{st}$ method are similar to the recent observations of spectroscopically confirmed LBGs 
at $z$=7.008 and $z=$7.109,
with line fluxes 1.62 $\times10^{-17}$ and 1.21 $\times  10^{-17}$ \ergscm, respectively \citep{van11}.
These line fluxes are accessible to deep spectroscopy with $>$8m telescopes, and thus these two candidates  provide good targets
for future spectroscopic followup.

\subsection{Specific Star Formation Rate}

The specific star formation rate, sSFR (SFR/$M_{\star}$) provides a way to quantify the evolution of mass build-up given a 
certain SFR.
Many observations suggest that the  sSFR is nearly constant from $z\sim7$ to $z\sim4$ \citep{sta09,gon10,bou11}.
This implies that the SFR increase from high-redshifts to low-redshifts, in conflict with the general assumption of 
constant or declining SFH. 
However, recent studies \citep[e.g.,][]{sta12,scha12} suggest that there is a  stronger evolution
in the sSFR from low-redshift to high-redshifts, than suggested by previous studies. Thus, to resolve this uncertainty among 
different studies,  large samples of high-redshift galaxies are needed.

Here we estimate the sSFR for the two brightest galaxies in our \zs\  sample  using the more accurate physical 
parameters derived from the fits to the \FS \mb + HST + IRAC data. 
We derived the SFRs using the best-fit SEDs: for the brightest LBG (zF13965), the  median SFR=77.6$^{+62}_{-42}$ \msunyear, much smaller
compared with the 2nd brightest LBG (zF25035) with median SFR=285.7$^{+346}_{-179}$ \msunyear (Table 5).

%The SFR can be obtained  using the Kennicutt relation  \citep{ken98}:
% \begin{equation}
%\rm SFR (M_{\odot} yr^{-1})=1.2\times 10^{28}  \; L_{\nu} \;\;erg \; s^{-1} Hz^{-1},
%SFR(\msun yr^{-1})=1.2\times 10^{28} \; L_{\nu} erg \; s^{-1} Hz^{-1}
%\end{equation} 
%
%where $L_{\nu}$ is the luminosity obtained from the best-fit SED at rest-frame 1500\AA.  
%The median stellar mass (Table 3) derived using \FS+\hst+IRAC  yield sSFR of 3.65 and 1.02 (Gyr$^{-1}$) for the two brightest
 %objects.  In Figure~14 we show the sSFR derived for the two brightest candidates (blue filled circles).

Our derived sSFR   $\sim$ 13 ~Gyr$^{-1}$ for the two  brightest \zs\  LBG candidate is  slightly higher compared  with 
 the  values derived for lower luminosity (and lower mass) galaxies at \zs, which 
implies a nearly constant sSFR from $z\sim7$ to $z\sim4$ \citep[e.g.,][]{red12} for galaxies over  approximately a decade in stellar mass 
($2\times 10^9$~\msun\ to $2\times 10^{10}$~\msun).    
%For the second brightest candidate, the sSFR is slightly lower, however with large uncertainty.
These sSFRs favors the idea that galaxies at these epochs have rising star-formation histories such that the sSFR is nearly independent 
of mass   at a fixed redshift   \citep[e.g.][]{pap11}.
  Models with constant or declining SFHs would  imply  decreasing sSFR with increasing mass at fixed redshift, which 
  is disfavored.   
  
  \begin{figure}[t!]

\epsscale{1.22}
\plotone{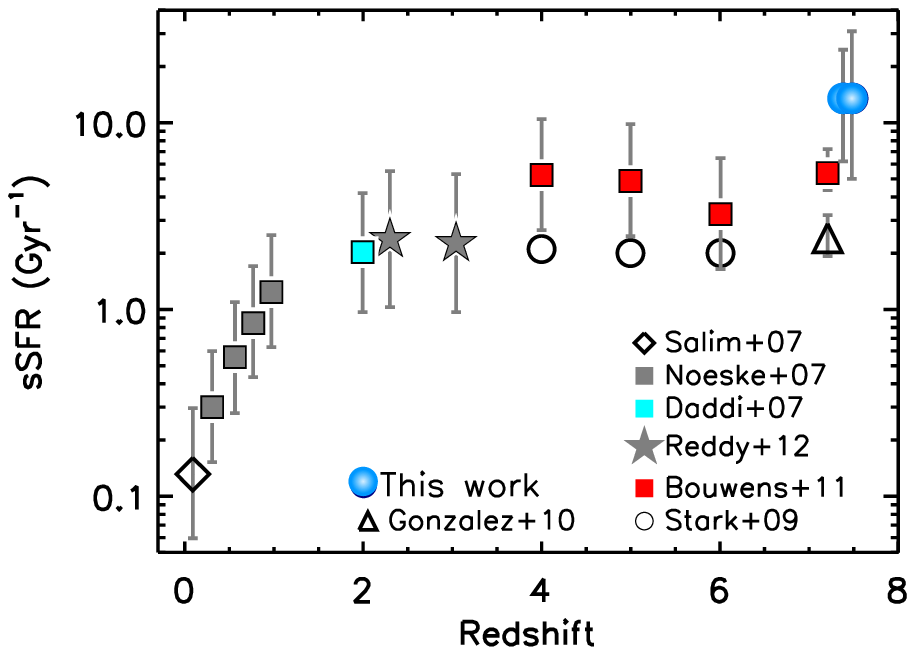}
\caption{ 
The  redshift evolution of the specific SFR.  Blue filled circles represent the  two brightest galaxies from \FS observations.
Open diamond, open circles, and open triangle represent data from \citet{sal07}, \citet{sta09}, and \citet{gon10} respectively
while  filled symbols- grey squares, cyan square, stars, and red squares show observations from \citet{noe07}, \citet{dad07}, 
\citet{red12}, and \citet{bou11} respectively.
% while
%filled squares, open circles, and open triangle represent data from Bouwens et al. (2011a), Stark et al (2009), and Gonzalez et al (2011)
%respectively. 
These galaxies have  stellar masses $\rm M_\star=(1-2) \times10^{10}$ \msun. The error bars on the \FS observations are the 
uncertainties in the estimated stellar mass obtained from the best-fit SED.
The sSFR remains nearly constant over 1 Gyr period from $z=8$ to
$z=4$,   which favors the idea of rising star-formation history with increasing stellar mass \citep{pap11}.\\
%Furthermore, our data show that the sSFR is nearly constant as
%a function of stellar mass at fixed redshift, which favors the idea of
%rising star-formation history i.e.  the SFR increases along with the
%the stellar mass from high-z to low-z (Papovich et al 2011).
%Star-formation histories that are constant or declining in time would
%favor sSFR that decrease with increasing mass at fixed redshift. \\
%\vspace{0.2cm}
}
\end{figure}

  \begin{figure}[t!]

\epsscale{1.22}
\plotone{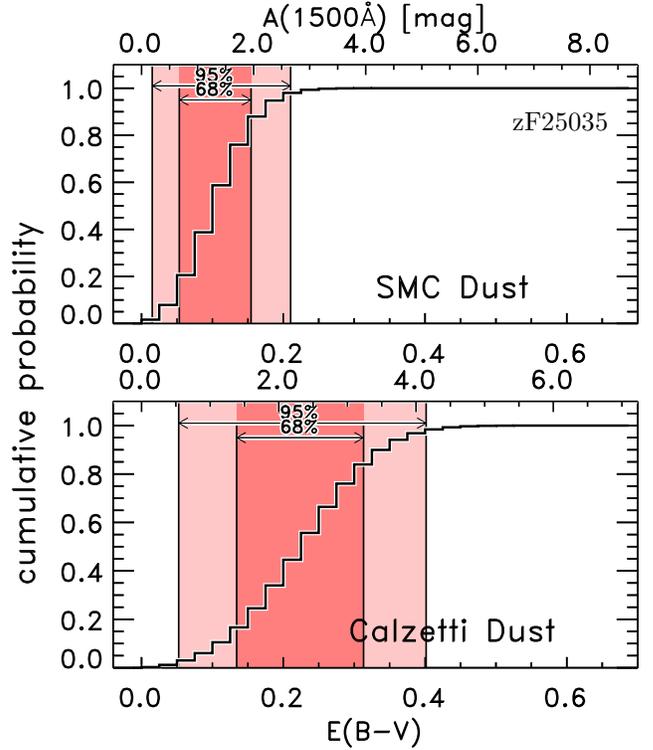}
{\normalsize\put(-60,235){zF25035}}
%{\normalsize \put(-60,105){zF25035}}
\caption{ 
Comparison of dust extinction derived from the best-fit SED using Calzetti and SMC dust extinction law.
Here we have shown the dust extinction of  the 2nd brightest LBG (zF25035) as this object yields significantly
different  extinctions for  SMC and  Calzetti dust extinction law, compared with the brightest LBG (zF13965).
The SMC-like law yields a median E(B-V)=0.1, a value much smaller compared with median E(B-V)=0.22 
obtained using the Calzetti dust extinction law.
}
\end{figure}

 \begin{table}[t!]
  \centering
  \caption{Comparison of  derived physical parameters from the best-fit SEDs using  Calzetti and SMC dust extinction laws for the
 two brightest LBGs.
  }
  \begin{threeparttable}
  \scriptsize{ 
      \begin{tabular}{lrr}
\hline

							&		zF13965								&		zF25035				\\
\hline

$\chi^2/\nu$					& 2.3		\; [1.8]								& 0.7 \;  [0.7]					\\[1.5ex]
Log Mass\scriptsize{(\msun) }		& 9.8$^{+0.2}_{-0.2}$ \;  [9.6$^{+0.3}_{-0.4}$]			& 10.3$^{+0.3}_{-0.4}$ \;  [9.8$^{+0.3}_{-0.4}$] \\[1.5ex]

E(B-V)						& 0.09$^{+0.06}_{-0.05}$ \;  [0.05$^{+0.03}_{-0.03}$]	& 0.22$^{+0.09}_{-0.09}$ \;  [0.1$^{+0.05}_{-0.05}$]									\\[1.5ex]
		
SFR\scriptsize {(\msun yr$^{-1}$)}	& 77.6$^{+62.3}_{-42.5}$	\;  [57.8$^{+32.7}_{-24.4}$]	& 285.7$^{+346.6}_{-179.6}$ [62.7$^{+75.7}_{-33.6}$]	\\ [1.5ex]

 sSFR\scriptsize {(Gyr$^{-1}$)}		& 13.5 $^{+1.8}_{-2.2}$ [13.5 $^{+2.0}_{-2.7}$]			& 13.5 $^{+2.3}_{-2.7}$ [10.7 $^{+2.3}_{-2.6}$]		
 \\[1.0ex]
 
  \hline   

        \end{tabular}}
    \begin{tablenotes}
      \scriptsize { 
       \item[] Values in square brackets are derived using SMC extinction law.
        }
          \vspace{0.5cm}
     \end{tablenotes}
        
  \end{threeparttable}

\end{table}

\subsection{Comparison Using Calzetti \& SMC Dust Extinction Laws}
To compare the effect of different extinction laws on the derived physical properties of \zs\  LBGs, 
we derived the best-fit SEDs (see Section 5.1.2) and physical properties of the two brightest \zs\  LBG candidates using
Calzetti and SMC dust extinction laws.

For the brightest LBG (zF13965), all derived parameters (stellar mass, age, and  $\chi^2$) using SMC dust extinction law
decrease except for the dust extinction (see Table 5). 
For the 2nd brightest LBG (zF25035), stellar mass, extinction and  $\chi^2$ either decrease or remain unchanged except
for the stellar age. A major difference among these two LBGs is that the dust extinction for zF25035 is significantly
reduced ($\rm A_{UV}$=0.9) when using the SMC extinction law (see Fig. 15). This smaller extinction value also gives rise to the lower
SFR compared with the Calzetti dust extinction-derived SFR.
%This demonstrates that using SMC dust extinction law (or  comparing it with the Calzetti law)
%is more appropriate for high-redshift LBGs, at least in some cases.
This demonstrates that the assumption of the extinction law affects the interpretation of the dust content in high-redshift galaxies.  
In particular, the SMC-type law - which may be more applicable in these distant galaxies \citep{oes12b} - reduces the
 implied dust content compared to a Calzetti-type law for low-redshift luminous starburst galaxies.

\section{Summary \& Conclusions}

We have obtained  a sample of \zs\   galaxies  from a deep  \nir  survey (\zf ) using   \mb  filters for the first time.  
We define a color-color selection criteria for the \nir  \mb filters, which cleanly isolate \zs\  LBG candidates  from brown dwarf stars and
low-redshift galaxies.  
Using our criteria we have identified  three robust  \zs\   candidate LBGs in a survey area of about 155 arcmin$^2$ in the 
COSMOS field, which has very deep publically available data.
The availability  of \nir   \mb   and  \bb   photometry allowed  us to compare the derived physical properties of high-redshift 
galaxies and how these properties are influenced by the choice of filter widths.  Our principal findings are summarized below.

\textit{Stars vs compact galaxies$-$}  
While the contamination from nearby \td  stars  is a serious concern for dropout-selected  galaxies, especially when using 
the ground-based \bb  photometry,  the \FS  \mb   filters allow us to cleanly distinguish between nearby dwarf stars and 
high-redshift galaxies that are relatively  compact.  

One of the three \zs\  LBGs  in our sample  (zF25035) is very compact and indistinguishable  from a point source at 
ground-based resolution (and  possibly unresolved even by HST).    
Our ability to distinguish this object from cool stars is possible only due to the  availability of  multiwavelength  \mb   
photometry.
Moreover, this demonstrates that there exists a population of very compact galaxies at \zs.

\textit{Bright end of the \zs\  UV luminosty function$-$} 
Using the number density of \zs\  galaxies  obtained using dropout
technique  and  corrected for incompleteness,
% using the selection
%function p(m,z), we derived 
the rest-frame stepwise UV luminosity  function at
\zs\  shows a moderate   evolution
from  $z\sim5$.  The number density of bright LBGs (\muv$\sim$-21.5)
increases by a factor of 4 from \zs\  to $z\sim 5$ albeit with large error bars.   
This is consistent with other ground-based studies but implies lesser  evolution compared
to the factor of 10 increase implied by some \hst-based studies.

\textit{Physical properties of \zs\  galaxies$-$}  
%We derived the
%physical properties of  two of our brightest galaxies at \zs\  based on
%the Bruzual \& Charlot models, and compared these derived properties
%with and without the \FS  MB  photometry. 
In
general, the presence of  \mb   photometry yield  tighter
constraints on the photometric redshifts and improved physical
constraints on the stellar population parameters of these galaxies. 

 \begin{enumerate}
 \setlength{\itemsep}{0pt}
  \setlength{\parskip}{0pt}
  \setlength{\parsep}{0pt}

   \item [(i)]For our two brightest LBGs, the best-fit SED model with \bb+ \mb  photometry   yields lower  photometric
   redshifts, and  much tighter redshift probability distribution compared to the \zphot$\sim$8 derived using only \bb  photometry
    (Figure~ 10). 
    This is also true for our 2nd brightest object.

\item[(ii)] For both objects, the UV spectral slope $\beta$  tends to be shallower  when derived using \bb  
photometry alone.  For the brightest object, $\beta$ changes from -1.88 to -2.08  when \FS \mb photometry is included.

  \item [(ii)] The SED modeling prefers a lower dust attenuation  with median E(B-V)=$0.01$ for \bb  photometry, while it 
  increases    to median E(B-V)=$0.09$ when including \mb photometry.  
  The \FS   bands allow for redder UV colors which increases the range of   permitted values for E(B-V). 
Comparing the above values, derived using the Calzetti dust law, with the E(B-V) derived using SMC-like dust law, we find that
the probability distribution from  the latter case is much narrower and  shifts towards lower values (Figure 15).

    \item [(iii)] Including the \FS \mb filters,  the SED modeling yields narrower probability distribution for stellar mass 
     for the two most luminous objects in our 
    sample. 
    We argue that adding additional bands shifts the allowed regions of the stellar-population parameter space, improving our 
    physical understanding of the galaxies. We find that the presence of IRAC photometry is also important to better constrain
    the best-fit stellar masses.

 \item [(iv)] We predict the \lya\ equivalent widths and line fluxes of the two brightest candidates and find that the 
 presence of \lya\ line influences the  derived physical properties of \zs\  candidates.
 The estimated rest-frame equivalent widths (17 and  26 \AA) make these two objects good targets for future spectroscopic followup.
 
\item[(v)] The  sSFRs for the two most luminous galaxies  are slightly larger compared with other \textit{lower mass}
galaxies at \zs. This favors the ideal that the SFR increases with increasing stellar mass  at this redshift.
%are consistent with  sSFRs for \textit{lower mass} 
%galaxies at \zs.   
%If larger samples confirm this trend, then this favors the idea that galaxies at these epochs have rising  star-formation histories  such that the specific 
%SFR is nearly  independent of mass at fixed redshift.   
%
%This implies that there is no measurable change in sSFR with mass at fixed redshift.   
%This favors the idea that galaxies at these epochs have rising  SFHs such that the specific SFR is nearly
 %independent of mass at fixed redshift.   
% Models with constant or declining SFHs would favor models of decreasing sSFR with increasing mass at fixed redshift, which 
 %is disfavored. 

\end{enumerate}

While it is early to reach   robust conclusions about the effect of \mb  photometry on the derived properties of \zs\  galaxies,
\mb  filters will likely provide better  constraints on the physical properties, due to their narrower redshift probability
distributions, compared to the \bb photometry.  
If this is indeed true, in the absence of spectroscopic redshifts,  multiwavelength \mb  photometry will provide the best 
constraints on the physical properties of \zs\  galaxies.

 \acknowledgments

This work was supported by the National Science Foundation grant
AST-1009707.  This work is based on observations taken by the CANDELS
Multi- Cycle Treasury Program with the NASA/ESA HST, which is operated
by the Association of Universities for Research in Astronomy, Inc.,
under NASA contract NAS5-26555. This work is supported by HST program
number GO-12060. Support for Program number GO-12060 was provided by
NASA through a grant from the Space Telescope Science Institute, which
is operated by the Association of Universities for Research in
Astronomy, Inc., under NASA contract NAS5-26555.   Australian access
to the Magellan Telescopes was supported through the National
Collaborative Research Infrastructure Strategy of the Australian
Federal Government.  LRS acknowledges funding from a Australian
Research Council Discovery Program grant DP1094370.  
This work is based in part on observations made with the Spitzer Space Telescope, 
which is operated by the Jet Propulsion Laboratory, California Institute of Technology under a contract with NASA. 
Support for this work was provided by NASA through an award issued by JPL/Caltech.
We acknowledge
generous support from the Texas A\&M University and the George P. and
Cynthia Woods Institute for Fundamental Physics and Astronomy.

%\todo{Casey added some references, make sure they end up in the table :)}

%\clearpage

%\begin{figure*}
%\epsscale{1.2}
%\plotone{sed_pz7.eps}
%\caption{Probability distribution of color selected objects that satisfy \zs\  dropout criteria. These p(z) are from EAZY.
%The probability $P(z>6)$ and the object IDs are shown as legends.  In addition to the color-selection criteria, only  objects with $P(z>6) > 0.7$  
% are selected as \zs\  galaxies.
%}
%\end{figure*}

%\begin{figure*}
%\epsscale{1.2}
%\plotone{sed_plot.eps}
%\caption{Spectral energy distributions of  objects that passed the color-selection criteria. These SEDs from from EAZY and makes use of all the
%available photometric data.
%}
%\end{figure*}

%\begin{figure*}
%\epsscale{1.2}
%\plotone{test_burg.eps}
%\caption{Best-fit observed spectra of M, L, and T-dwarfs to medium-bands of color-selected \zs\  objects. The medium-bands are most powerful in 
%identifying T-dwarfs  (IDs 12981, 12518).  M \& L dwarfs are little bit harder to discriminate due to their flat spectra resembling galaxy-spectra.
%}
%\end{figure*}

\end{document}